\documentclass[aps,preprint, showkeys]{revtex4}
\usepackage{amsmath}
\usepackage{graphicx}
\newcommand{\be}{\begin{equation}}
\newcommand{\ee}{\end{equation}}
\newcommand{\bea}{\begin{eqnarray}}
\newcommand{\eea}{\end{eqnarray}} 
\newcommand{\ba}{\begin{array}}
\newcommand{\ea}{\end{array}}

\newcommand{\bb}{\bibitem}
\begin{document}

\title{The simplest minimal subtraction for massive scalar field theory}
\author{Marcelo M. Leite\footnote{email: {marcelo.mleite@ufpe.br}}}
\affiliation{Laborat\'orio de F\'\i sica Te\'orica e Computacional, Departamento de F\'\i sica,\\ Universidade Federal de Pernambuco,\\
50670-901, Recife, PE, Brazil}
\begin{abstract}
 {\it The simplest minimal subtraction method for massive $\lambda \phi^{4}$ scalar field theory is presented.  We utilize the one-particle irreducible vertex parts framework to deal only with the primitive divergent ones that can be renormalized multiplicatively.  We give a unified description for spacetime metric tensor with either Minkowski or Euclidean signature.The partial-$p$ operation in the remaining diagrams of the two-point vertex part eliminates its overlapping divergences.   We show how the parametric dissociation transform effectively removes the external momentum dependence of the coefficient of the squared bare mass after performing the  partial-$p$ operation in the two-point vertex part diagrams.  The resemblance of this method with a minimal subtraction scheme in the massless theory is pointed out. We derive the Callan-Symanzik equations using minimal subtraction arguments and discuss the scaling limit in the ultraviolet region. We apply the method to determine critical exponents for an $O(N)$ internal symmetry at least up to two-loop order with a flat Euclidean metric and find perfect agreement with all previous results in the literature.}   
\end{abstract}

\maketitle

\section{Introduction}
\par Massive scalar field theories are worth studying for at least two reasons. First, its scaling at the ultraviolet regime \cite{BLZ} makes a nontrivial connection between critical phenomena \cite{WK}, relativistic field theory \cite{GL} and its applications in particle physics \cite{H}.  In this form, the renormalization group was utilized to extract finite outcomes from otherwise infinite meaningless results in quantum field theory ($QFT$). Its association with critical phenomena manifested through the ultraviolet momentum cutoff \cite{Wilson1,Wilson2,WK}.  Second,  the Callan-Symanzik ($CS$) equations \cite{CS,C1,C2} which govern the behavior of the solution of the one-particle irreducible ($1PI$) vertex parts for massive fields under scaling were obtained so far only through the utilization of normalization conditions, although not making explicit reference to a cutoff.  In the scaling regime, it yields identical results as the renormalization group equations for massless fields\cite{BLZ1}.  Normalization conditions fix the external momenta scale of certain physical quantities from the outset in finite values (differing by finite amounts of their on-shell values) for the renormalized theory.  They are convenient for studying low-energy (near mass shell) behavior. 
\par  However, the description of high-energy behavior as in deep inelastic scattering \cite{G} is typically an off-shell effect, which requires taking the limit of the external momenta (Euclidean or spacelike in the case of Minkowski space) to very large values \cite{W1}.  Minimal subtraction \cite{tHV,tH,W2,LA} is the momentum-independent renormalization scheme which is more suited for studying high-energy behavior.  We could ask ourselves whether, starting from totally cutoff independent arguments, we could achieve the connection between 
$QFT$ and critical phenomena with perhaps additional information concerning ultraviolet and infrared behaviors. A derivation of the $CS$ equation in a mass-independent scheme such as minimal subtraction  along with the investigation of its properties in the ultraviolet regime would be undoubtedly desirable to give decisive arguments using only perturbation theory.  
\par This work is devoted to provide the details of the method briefly described in \cite{Leite1}. Namely, we present the modification of the diagrammatic expansions of the primitively divergent vertex parts and their renormalization resulting in the simplest minimal subtraction method for the massive scalar field theory to date, with a minimal number of diagrams. We can then understand why there is a considerable deviation in defining this scheme for the massless and massive fields.  Note that this is in stark contrast with normalization conditions where the renormalization of massive and massless cases is very similar. 
\par Indeed, this is a longstanding problem regarding the formal properties of minimal subtraction.  Suppose the problem is formulated without requiring a minimal number of diagrams in the perturbative expansion. In that case,  the $BPHZ$ method \cite{BPHZ,K} seems to be a good candidate to put the massless and massive theories on an equal footing.  However, the $CS$ equations as obtained in \cite{K} utilize normalization conditions  of the $1PI$ composed field vertex part  as $\Gamma_{R}^{(N,1)}=1$, which is the original argument in \cite{C2} using normalization conditions exclusively. An attempt to avoid this problem introduced more structure in the $CS$ equations without addressing the practical issue of implementing the argument order by order in the Feynman diagrams of the two-point vertex part \cite{NN}.  
\par This issue is resolved in the present work, with a minimal number of diagrams.  In particular, the simple form of the perturbative expansion of the $1PI$ vertex parts makes it unnecessary to delve into further steps concerning the integration by parts procedure to compute arbitrary Feynman integrals (contributing to scattering amplitudes in relativistic collisions; see, for example \cite{T,L,He}).
\par We explicitly explore the computation of Feynman integrals by circumventing the overlapping divergences in the two-point vertex part with a simple prescription.  The definition of the renormalized mass in terms of the bare mass and one of the normalization functions {\it at all orders in perturbation theory} is very similar to what occurs in the massless theory. It makes the renormalization automatic and extremely simple.  We describe the method using the spacetime metric tensor with Minkowski as well as Euclidean signature. 
\par In addition, we show how to obtain the $CS$ equations using the minimal subtraction method exclusively. Consequently,  the dissimilarity between massive and massless integrals in the context of normalization conditions and minor modifications (practically the same when applied to both settings) is extended to minimal subtraction, establishing a "covariance" with respect to different renormalization schemes. As applications, we analyze 
the ultraviolet regime in $QFT$ and compute by diagrammatic means the critical exponents $\eta$ and $\nu$ using a flat metric tensor with a Euclidean signature.
\par The organization of the paper is as follows: in Sec. II we briefly review how the tadpoles can be eliminated in the diagrammatic expression of the primitively divergent vertex part by a suitable redefinition of the bare mass at three-loop order.   Section III is devoted to showing how to handle the overlapping divergences simultaneously using  the partial-$p$ operation along with the $PDT$ technique.   Section IV presents the renormalization of the primitively divergent vertex parts.  In Sec. V the Callan-Symanzik equation is obtained without any reference to normalization conditions. We then investigate the ultraviolet behavior of the $\phi^{4}$ theory with Minkowski spacetime metric tensor as well as the Euclidean version adequate to address critical phenomena,  and show whether there is consistency of the 
fixed points with actual ultraviolet fixed points. The computation of the critical exponents  $\eta$ and $\nu$ by diagrammatic means as an application of the framework developed for  Euclidean spacetime metric is the subject of Sec. VI.  Section VII contains our conclusions. In 
Appendices A and B,  only the Feynman diagrams which survive the tadpole cancellations explained in the body of the paper are expressed in terms of integrals along with their $\epsilon$-expansions for the spacetime metric tensor with Minkowski and Euclidean signature.
\section{Brief review of tadpole cancellations}
\par In the present section, our discussion will be purely diagrammatic: we will not care about the precise expressions 
of the diagrams in terms of integrals but only describe the mechanism of tadpole cancellation.  The aim is to give a unified description of quantum field theory defined in Minkowski space and as an order parameter in the context of static critical phenomena in Euclidean space.  The precise expressions, including each diagram factor, and the appropriate coupling constant power can be found in Appendices A and B.
\par To begin with, we promote a slight change of notation in connection with our previous work \cite{Leite1}. The reason is that we want to unify the problem of a scalar quantum field on Minkowski spacetime and Euclidean space.  Consider the bare Lagrangian density for the scalar field with $O(N)$ internal symmetry, defined on a $d$-dimensional flat background. If the metric tensor has Minkowski (Euclidean) spacetime (space) signature
$(+,-,...,-)$ ($(+,...,+)$) with index 
$\nu=0,1,..., d-1 (\nu=1,..., d)$ , it can be written as
\begin{equation}\label{1}
\mathcal{L} = \frac{1}{2} \partial_{\nu} \phi \partial^{\nu} \phi - \frac{1}{2} \mu_{0}^{2} \phi^{2} 
- \frac{\lambda}{4!} (\phi^{2})^{2},
\end{equation} 
where $\mu_{0}$ and $\lambda$ are the bare mass and coupling constant, respectively. An implicit 
index in the scalar fields will be omitted throughout this work since they produce an $N$-dependent overall factor characterizing each diagram in perturbation theory.  The objects $\Gamma^{(2)}(p, \mu_{0}, \lambda, \Lambda)$, $\Gamma^{(4)}(p_{i}, \mu_{0}, \lambda, \Lambda)$ and $\Gamma^{(2,1)}(p_{1}, p_{2}; Q,\mu_{0}, \lambda, \Lambda)$ are the primitively divergent vertex parts. The $p_{i}$ in the argument of the several vertex parts stands for external momenta, whereas $Q$  is the momentum of the inserted composite operator,  respectively.  The parameter $\Lambda$ (the cutoff) will be implicit until explicitly necessary when we discuss the scale properties of the renormalized vertex parts.  
\par Before proceeding, let us discuss the differences between the Minkowski and Euclidean spaces.  For the former, the weight factor in the functional integral defining the generating functional is $exp[i \int dx^{0}dx^{1}...dx^{d-1} \mathcal{L}]$.  Propagators are given by $\frac{i}{p^{2} - \mu_{0}^{2}}$ (with $p^{2} = (p^{0})^{2} - (p^{1})^{2} -...-(p^{d-1})^{2}$),  whereas the contribution of the $\phi^{4}$ vertex in the diagrammatic expansion is proportional to a factor $(-i\lambda)^{n}$ at $n$th order in the coupling constant. The bare mass at a fixed order in the perturbative expansion will be defined by $\mu^{2} = -  \Gamma^{(2)}(p=0, \mu_{0})$. Within this nomenclature the two-particle scattering amplitude is defined by $i \mathcal{M}(p_{1}p_{2} \rightarrow p_{3} p_{4}) = \Gamma^{(4)}(p_{i}, \mu_{0}, \lambda)$ (see, for instance \cite{PS}).  The tree-level value of the composite field $\Gamma^{(2,1)}(p_{1}, p_{2}; Q,\mu_{0}, \lambda, \Lambda)$ is equal to one for (both) Minkowski (and Euclidean) signature(s) of the metric tensor.  In the diagrammatic expansion of $\Gamma^{(4)}(p_{i}, \mu_{0}, \lambda)$, 
we have to include the contributions of the $s$, $t$ and $u$ channels, where $s=(p_{1}+p_{2})^{2}$, $t=(p_{1}+p_{3})^{2}$, and $u=(p_{2}+p_{3})^{2}$ are the Mandelstam variables.    
\par The Euclidean metric signature has also some differences from our previous work. First the partition function has the weight factor obtained by Wick rotation of time $x^{0}=-ix^{d}$, namely $exp[\int dx^{1}...dx^{d} \mathcal{L}]$. This has the standard form since all the terms in the bare Lagrangian density are negative and the weight factor is an exponential of a negative quantity.With this new notation for the bare Lagrangian density,  $\Gamma^{(4)}(p_{i}, \mu_{0}, \lambda, \Lambda)$ in the present paper corresponds to $-\Gamma^{(4)}(p_{i}, \mu_{0}, \lambda, \Lambda)$ in ref. \cite{Leite1}. In addition, odd powers of the coupling constant receive a minus sign in all primitively divergent vertex parts (without the factor of $i$ present in the Minkowski case).  The free propagator is $\frac{1}{p^{2} + \mu_{0}^{2}}$ (with $p^{2} = (p^{1})^{2} + (p^{2})^{2} +...+(p^{d})^{2}$) and the bare mass at a given order in the perturbative expansion of the coupling constant is defined by $\mu^{2} =  \Gamma^{(2)}(p=0, \mu_{0})$.  
\par The idea of the elimination of tadpole diagrams in the perturbative expansion appeared first in the massless theory \cite{Amit}.  A brief review of a previous proposal for the massive theory which is efficient to get rid of tadpole insertions \cite{CarLei} at all loop orders is in order.  We shall restrict our discussion to the loop order convenient to our purposes.   
\par First, we write down the diagrammatic expansion of the primitively divergent vertex part $\Gamma^{(2)}(p, \mu_{0}, \lambda)$ in terms of the proper self-energy part ("mass operator") $\Sigma(p, \mu_{0},\lambda)$.  The definition of the latter depends whether we consider the Minkowski or Euclidean cases as previously mentioned. For Minkowski signature the definition we utilize is $\Gamma^{(2)}(p, \mu_{0}, \lambda)= p^{2} - \mu_{0}^{2} - \Sigma_{M}(p, \mu_{0},\lambda)$ \cite{Ry}, whereas in the Euclidean case $\Gamma^{(2)}(p, \mu_{0}, \lambda)= p^{2} + \mu_{0}^{2} - \Sigma_{E}(p, \mu_{0},\lambda)$, where up to three-loop order
\begin{subequations}
\begin{eqnarray}\label{2}
&& \Sigma_{M}(k, \mu_{0},\lambda)  =   i \Bigl[\parbox{12mm}{\includegraphics[scale=0.7]{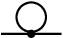}} +
\parbox{12mm}{\includegraphics[scale=0.7]{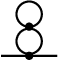}} + \;\parbox{12mm}{\includegraphics[scale=0.7]{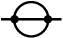}} +\; \parbox{12mm}{\includegraphics[scale=0.7]{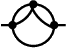}}\nonumber\\
&&  +\; \parbox{12mm}{\includegraphics[scale=0.7]{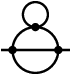}} +\; \parbox{12mm}{\includegraphics[scale=0.7]{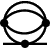}} + 
\;\parbox{12mm}{\includegraphics[scale=0.7]{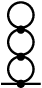}} + \;\parbox{12mm}{\includegraphics[scale=0.7]{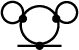}}\Bigr]_{M},\label{2a}\\
&& \Sigma_{E}(k, \mu_{0},\lambda)  =  \Bigl[ \parbox{12mm}{\includegraphics[scale=0.7]{fig1.eps}} +
\parbox{12mm}{\includegraphics[scale=0.7]{fig2.eps}} + \;\parbox{12mm}{\includegraphics[scale=0.7]{fig6.eps}} +\; \parbox{12mm}{\includegraphics[scale=0.7]{fig7.eps}}\nonumber\\
&&  +\; \parbox{12mm}{\includegraphics[scale=0.7]{fig8.eps}} +\; \parbox{12mm}{\includegraphics[scale=0.7]{fig5.eps}} + 
\;\parbox{12mm}{\includegraphics[scale=0.7]{fig3.eps}} + \;\parbox{12mm}{\includegraphics[scale=0.7]{fig4.eps}}\Bigr]_{E}.
\end{eqnarray}
\end{subequations}
\par The explicit subscripts refer to the computation of the diagrams utilizing the Feynman rules appropriate for each case. Consider the Euclidean version with our new conventions.  Make a change of the bare mass parameter. We replace its tree-level value $\mu_{0}$ by the 
three-loop bare mass parameter through the expression $\mu^{2}=\Gamma^{(2)}(k=0, \mu_{0},\lambda)$, or explicitly
\begin{eqnarray}\label{3}
&&\mu^2  =  \mu_{0}^{2} \; + \; \parbox{12mm}{\includegraphics[scale=0.7]{fig1.eps}} \; + \; \parbox{12mm}{\includegraphics[scale=0.7]{fig2.eps}} \; + \parbox{12mm}{\includegraphics[scale=0.7]{fig6.eps}}\bigg|_{k=0} \nonumber\\
&&  \;  + \;  \parbox{12mm}{\includegraphics[scale=0.7]{fig7.eps}}\bigg|_{k=0}  + \; \parbox{12mm}{\includegraphics[scale=0.7]{fig8.eps}}\bigg|_{k=0}  \nonumber\\
&& \;+ \; \parbox{12mm}{\includegraphics[scale=0.7]{fig5.eps}} \; + \;  
\parbox{12mm}{\includegraphics[scale=0.7]{fig3.eps}} \quad + \quad  
\parbox{12mm}{\includegraphics[scale=0.7]{fig4.eps}}. \label{3}
\end{eqnarray}
\par By writing $\mu_{0}$ as a function of $\mu$ through the inversion of the last equation,  the resulting
$\Gamma^{(2)}(k,\mu_{0}(\mu),\lambda)$ no longer possesses tadpole insertions. These graphs  will 
vanish into the $\Gamma^{(2)}(k,\mu_{0}(\mu),\lambda)$ expression obtained after the substitution 
$\mu_{0}(\mu)$.  Not only this, but the other primitively divergent vertex parts $\Gamma^{(4)}(p_{i}, \mu_{0}(\mu), \lambda)$ and  $\Gamma^{(2,1)}(p_{1}, p_{2}; Q,\mu_{0}(\mu), \lambda)$ turn out tadpole-free after this manipulation. Explicitly, we are left with a minimal number of diagrams for those vertex parts that now can be written solely in terms of $\mu$ as
\begin{eqnarray}\label{4}
&&\Gamma^{(2)} (p,\mu,\lambda) =  p^{2} + \mu^{2} \nonumber\\
&& - \left(\parbox{12mm}{\includegraphics[scale=0.7]{fig6.eps}}\bigg|_{\mu} \;- \;
\parbox{12mm}{\includegraphics[scale=0.7]{fig6.eps}}\bigg|_{p=0, \mu}\right) \nonumber \\ 
&&  - \;  \left(\parbox{12mm}{\includegraphics[scale=0.7]{fig7.eps}}\bigg|_{\mu}  \; - \;
\parbox{12mm}{\includegraphics[scale=0.7]{fig7.eps}}\bigg|_{p=0, \mu}\right) ,
\end{eqnarray}
\begin{eqnarray}\label{5}
&\Gamma^{(4)}(p_{i},\mu,\lambda) =  -\lambda + 
\Bigl(\Bigl[\parbox{12mm}{\includegraphics[scale=0.7]{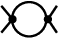}}(p_{1}+p_{2})\Bigr]_{\mu}\nonumber\\
&+ 2perms.\Bigr) + \Bigl(\Bigl[\parbox{12mm}{\includegraphics[scale=0.7]{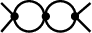}}(p_{1}+p_{2})\Bigr]_{\mu} + 2perms.\Bigr)\nonumber\\
& + \Bigl(\Bigl[\parbox{12mm}{\includegraphics[scale=0.7]{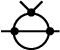}} \Bigr]_{\mu}(p_{i}) + 5perms. \Bigr),
\end{eqnarray}
\begin{eqnarray}\label{6}
&\Gamma^{(2,1)}(p_{1},p_{2};Q,\mu,\lambda) = 1 + \nonumber\\ 
&\Bigl(\Bigl[\parbox{12mm}{\includegraphics[scale=0.7]{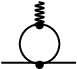}}\Bigr]_{\mu}(p_{1}+p_{2}) + 2perms.\Bigr)\nonumber\\
& + \Bigl(\Bigl[\parbox{12mm}{\includegraphics[scale=0.7]{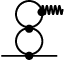}}\Bigr]_{\mu}(p_{1}+p_{2}) + 2perms.\Bigr)\;  +\nonumber\\
&+ \Bigr(\Bigr[\parbox{12mm}{\includegraphics[scale=0.7]{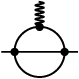}}\Bigr]_{\mu}(p_{1},p_{2};Q)  + 5perms.\Bigr). 
\end{eqnarray}   
\par In the Minkowski case,  the above modified diagrammatic expansion including the thee-loop bare mass has the same general form.  The minor modifications come from the explicit expressions of the propagator and coupling constant factors in the Feynman diagrams.  For instance, the two-point function after the replacement of $\mu_{0}(\mu)$ along the same lines just discussed yields
\begin{eqnarray}\label{7}
&&\Gamma^{(2)} (p,\mu,\lambda) =  p^{2} - \mu^{2} \nonumber\\
&& - i \left(\parbox{12mm}{\includegraphics[scale=0.7]{fig6.eps}}\bigg|_{\mu} \;- \;
\parbox{12mm}{\includegraphics[scale=0.7]{fig6.eps}}\bigg|_{p=0, \mu}\right) \nonumber \\ 
&&  -i \;  \left(\parbox{12mm}{\includegraphics[scale=0.7]{fig7.eps}}\bigg|_{\mu}  \; - \;
\parbox{12mm}{\includegraphics[scale=0.7]{fig7.eps}}\bigg|_{p=0, \mu}\right) ,
\end{eqnarray}
\par Afterward, we do not need to worry about the integrals including tadpole insertions of these primitively divergent vertex parts that were present in the first place but vanished using this technique.   Appendices A and B should be consulted to associate each Feynman diagram with the corresponding momentum space integral.  There will be a short discussion of the $\epsilon$-expansion of the integrals involved in the vertex parts $\Gamma^{(2)}$ and $\Gamma^{(4)}$. Note that the diagrams from $\Gamma^{(2,1)}$ differ from those from $\Gamma^{(4)}$ only by a global factor and can be easily reconstructed from our discussion in the Appendices; see Sec. IV.  
\section{The parametric dissociative transform}
\par The main problem for a successful minimal subtraction scheme is the treatment of the two- and three-loop integrals corresponding to the two-point function's second and third terms of the diagrammatic expansion in 
Eqs.  (\ref{4}) and (\ref{7}).   
\par A first attempt to do this employing only the partial-$p$ operation in those diagrams led to a finite answer to  
$\Gamma^{(4)}$ and $\Gamma^{(2,1)}$. Still,  the two-point function retained a residual single pole depending on the bare mass parameter $\mu$.  In other words, the minimal subtraction procedure did not work for $\Gamma^{(2)}$.  To remedy this undesirable aspect, the vertex function was redefined with an extra subtraction of the simple pole {\it after} setting the minimal subtraction \cite{CarLei}.  This unsatisfactory state of affairs has consequences on the theory's nonperturbative properties, as shown in Sec. V.  We now describe how to circumvent this difficulty to set 
a straightforward minimal subtraction scheme.
\par Although the arguments are entirely analogous to Minkowski and Euclidean metric signatures,  we now outline 
these different versions separately to avoid any possibility of misunderstanding in the unified description of this technique.  We will be a little more explicit in discussing the situation with the Euclidean metric signature as a follow-up discussion to reference \cite{Leite1}.  
\subsection{Minkowski signature}
\par Consider the two-loop diagram of the vertex part $\Gamma^{(2)}$, whose associated integral is given by
\begin{eqnarray}\label{8}
&I_{3}(p,\mu) = \int \frac{d^{d} q_{1} d^{d} q_{2}}{(q_{1}^{2} - \mu^{2})(q_{2}^{2} - \mu^{2})[(q_{1} + q_{2} + p)^{2} - \mu^{2}]}.
\end{eqnarray}
We apply the partial-$p$ operation in the form $\frac{1}{2d} [\frac{\partial q_{1}^{\mu}}{\partial q_{1}^{\mu}} 
+ \frac{\partial q_{2}^{\mu}}{\partial q_{2}^{\mu}}]$, where $q_{i}$ ($i=1,2$) are the loop momentum.  We then obtain
\begin{subequations}\label{9}
\begin{eqnarray}
&I_{3}(p,\mu) = \frac{1}{(d-3)} \Bigl[3 \mu^ {2} A_{M}(p,\mu)- B_{M}(p,\mu) \Bigr], \label{9a}\\
&A_{M}(p,\mu) = \int  \frac{d^{d} q_{1} d^{d} q_{2}}{(q_{1}^{2} - \mu^{2})^{2} (q_{2}^{2} - \mu^{2})[(q_{1} + q_{2} + p)^{2} - \mu^{2}]}, \label{9b}\\
&B_{M}(p,\mu) = \int  \frac{d^{d} q_{1} d^{d} q_{2}  p. (q_{1} +q_{2} +p)}{(q_{1}^{2} - \mu^{2}) 
(q_{2}^{2} - \mu^{2})[(q_{1} + q_{2} + p)^{2} - \mu^{2}]^{2}}.\label{9c}
\end{eqnarray}
\end{subequations}
\par When we compare this massive integral with that from the massless theory, there are two essential points. The first one is that we do not need to carry out the partial-$p$ operation in the massless integral, since the integral we started with can be solved.  Second,  the renormalized mass keeps its vanishing bare value.  
\par We have several possibilities to attempt to implement a simple prescription to solve the massive integral along the same lines as the massless theory.  Setting $\mu=0$ in the coefficient of the integral $A(p,\mu)$ is ruled out for a massive theory.  Another possibility is to set $p=0$ inside $A(p,\mu)$ but this is rather artificial,  since we still have to subtract the value of the diagram at $p=0$.  A more satisfactory solution is to proceed and try to identify an intermediate step which gets rid of the external momentum dependence of the integral $A_{M}(p,\mu)$.  If we can do that for all integrals which appear in arbitrary loop order for the vertex part  $\Gamma^{(2)}$,  
then a final step would be to fix the renormalized mass without referring to the renormalization of the primitively divergent vertex parts.
\par The implementation of the parametric dissociation transform ($PDT$) \cite{Leite1} is described in Appendix A.  We just highlight the idea here as follows.  For $I_{3}$ we solve the subdiagram and employ an additional Feynman parameter to obtain the solution of the remaining loop momentum integral.The latter is solved with a particular set of values of the Feynman parameters which eliminates the external momentum dependence.  
\par By employing the formulae obtained in Appendix A, we find
\begin{eqnarray}\label{10}
&I_{3}(p,\mu)|_{PDT} = -\frac{3(\mu^{2})^{1-\epsilon}}{2\epsilon^{2}} \Bigl[1+\frac{1}{2} \epsilon + \epsilon^{2}\Bigl(\frac{\pi^{2}}{12} + 1 \Bigr)\Bigr] \nonumber\\
& +\frac{p^{2}(\mu^{2})^{-\epsilon}}{8\epsilon} \Bigl[1+\frac{1}{4} \epsilon  - 2\epsilon L_{3M}(p,\mu)\Bigr].
\end{eqnarray}
\par The first term comes from the integral $A_{M}(p,\mu)$ using the $PDT$,  whereas the second term comes from the integral $B_{M}(p,\mu)$ and does not require any particular value of any Feynman parameters.  
\par The integral associated with the three-loop diagram of the two-point function is given by
\begin{eqnarray}\label{11}
&I_{5}(p,\mu) = \int \frac{d^{d} q_{1} d^{d} q_{2} d^ {d} q_{3}}{(q_{1}^{2} - \mu^{2})(q_{2}^{2} - \mu^{2})(q_{3}^2 - \mu^ {2})[(q_{1} + q_{2} + p)^{2} - \mu^{2}]}\nonumber\\
& \times \; \frac{1}{[(q_{1} + q_{3} + p)^{2} - \mu^{2}]}.
\end{eqnarray}
\par The appplication of the partial-$p$ operation in this integral yields
\begin{subequations}\label{12}
\begin{eqnarray}
&I_{5}(p) = \frac{2}{(3d-10)} \Bigl[\mu^{2} (C_{1M}(p,\mu) + 4 C_{2M}(p,\mu)) \nonumber\\
& \qquad - 2 D_{M}(p,\mu)\Bigr],\label{12a}\\
&C_{1M}(p,\mu) = \int \frac{d^{d} q_{1} d^{d} q_{2} d^ {d} q_{3}}{(q_{1}^{2} - \mu^{2})^{2}(q_{2}^{2} - \mu^{2})(q_{3}^2 - \mu^ {2})[(q_{1} + q_{2} + p)^{2} - \mu^{2}]}\nonumber\\
& \times \; \frac{1}{[(q_{1} + q_{3} + p)^{2} - \mu^{2}]},\label{12b}\\
&C_{2M}(p,\mu) = \int \frac{d^{d} q_{1} d^{d} q_{2} d^ {d} q_{3}}{(q_{1}^{2} - \mu^{2})(q_{2}^{2} - \mu^{2})^{2} (q_{3}^2 - \mu^ {2})[(q_{1} + q_{2} + p)^{2} - \mu^{2}]}\nonumber\\
& \times \; \frac{1}{[(q_{1} + q_{3} + p)^{2} - \mu^{2}]},\label{12c}\\
&D_{M}(p,\mu) =  \int  \frac{d^{d} q_{1} d^{d} q_{2} d^{d} q_{3} p.(q_{1} + q_{2} + p)}{(q_{1}^{2} - \mu^{2}) (q_{2}^{2} - \mu^{2})(q_{3}^{2} - \mu^{2})} \nonumber\\
&\qquad \frac{1}{[(q_{1} + q_{2} + p)^{2} - \mu^{2}]^{2}[(q_{1} + q_{3} + p)^{2} - \mu^{2}]}.\label{12d}
\end{eqnarray}
\end{subequations}
\par Although we can obtain the result above by straightforward manipulations, we can atribute a deeper meaning to the appearance of two different integrals $C_{1M}(p,\mu)$,  $C_{2M}(p,\mu)$ after the partial-$p$ operation using homotopy topological arguments \cite{AP}.  The reader can find other perturbative approaches for numerical computations and involving topological arguments in \cite{AV}.
\par Let us return for a moment and look at $A_{M}(p,\mu)$ of the two-loop contribution.  We can understand the factor of 3 in front of $A_{M}(p,\mu)$ by the topological property of this particular diagram.  All loops can be contracted to a point equally in the three propagators hit by the partial-$p$ operation resulting in the same contracted diagram corresponding to a tadpole.  
\par We can also employ arguments of homotopy theory to understand the different coefficients in the integrals 
$C_{1M}(p,\mu)$ and $C_{2M}(p,\mu)$.  Indeed,  the partial-$p$ operation hits four propagators of the two 
one-loop four-point vertex part $\Gamma^{(4)}$ subdiagrams that produce $C_{2M}(p,\mu)$.  When the loop containing any propagator of this type is contracted to a point,  we obtain 4 topologically identical diagrams which are nothing but the two-loop contribution just discussed. On the other hand, the partial-$p$ operation on the top propagator of the original diagram will produce  $C_{1M}(p,\mu)$.  The contraction of this loop diagram produces a nontrivial diagram that cannot be identified with the two-loop contribution as it were for the integral 
$C_{2M}(p,\mu)$.  This general argument proves why these two integrals are different as well as their $\epsilon$-expansions. 
\par Now the $PDT$ implementation on $C_{1M}(p,\mu)$ integral corresponds to the choice of endpoint singularities of the intermediate parametric integrals to solve the remaining momentum integral (after the solution of the subdiagrams). These fixed values of parameters do not correspond to endpoint singularities of the integral $C_{2M}(p,\mu)$ owing to its different homotopy properties compared with $C_{1M}(p,\mu)$. Nevertheless, the same set of parameters in both cases makes them independent of the external momentum when the last momentum integral is performed at these fixed sets of Feynman parameters.
\par As explained in Appendix A, the manipulations appropriate to Minkowski signature leads to the following result:
\begin{eqnarray}\label{13}
&I_{5}(p,\mu)|_{PDT} = \frac{i \mu^{2-3\epsilon}}{\epsilon^{3}}\Bigl[ 1-  \frac{\epsilon}{3} + 
\frac{19}{12}\epsilon^{2} + \frac{ 3 \psi'(1)}{4}\epsilon^{2}\Bigr] \nonumber\\
&+ \frac{i p^{2} \mu^{-3 \epsilon}}{6 \epsilon^{2}}\Bigl[1 + \frac{\epsilon}{2} -3 \epsilon L_{3M}(p,\mu) \Bigr].
\end{eqnarray}   
\par An arbitrary loop graph (without tadpoles) will appear in the perturbative expansion of the vertex part  
$\Gamma^{(2)}$ always subtracted from its value of $p=0$ in this setting.  After applying the partial-$p$ operation, the coefficient of $\mu^{2}$ is composed, say, by a set of homotopically different $\mathcal{C}_{iM} (p,\mu)$ auxiliary integrals in a sense just discussed for the two- and three-loop cases. They can be made momentum-independent with the same fixed Feynman parameters in the last loop momentum integral in any order in perturbation theory. This argument can be easily extended to the Euclidean signature as well.
\subsection{Euclidean signature}
We apply the partial-$p$ operation in the form $\frac{1}{2d} [\frac{\partial q_{1}^{\mu}}{\partial q_{1}^{\mu}} 
+ \frac{\partial q_{2}^{\mu}}{\partial q_{2}^{\mu}}]$, with the Euclidean metric just as before.  We then obtain
\begin{subequations}\label{14}
\begin{eqnarray}
&I_{3}(p,\mu) = -\frac{1}{(d-3)} \Bigl[3 \mu^ {2} A_{E}(p,\mu)+ B_{E}(p,\mu) \Bigr], \label{14a}\\
&A_{E}(p,\mu) = \int  \frac{d^{d} q_{1} d^{d} q_{2}}{(q_{1}^{2} + \mu^{2})^{2} (q_{2}^{2} + \mu^{2})[(q_{1} + q_{2} + p)^{2} + \mu^{2}]}, \label{14b}\\
&B_{E}(p,\mu) = \int  \frac{d^{d} q_{1} d^{d} q_{2}  p. (q_{1} +q_{2} +p)}{(q_{1}^{2} + \mu^{2}) 
(q_{2}^{2} + \mu^{2})[(q_{1} + q_{2} + p)^{2} + \mu^{2}]^{2}}.\label{14c}
\end{eqnarray}
\end{subequations}
We apply $PDT$ as explained in Appendix B and find
\begin{eqnarray}\label{15}
&& A_{E}(p,\mu)|_{PDT}= \frac{(\mu^{2})^{-\epsilon}}{2 \epsilon^{2}}\Bigl[1 -\frac{\epsilon}{2} + \epsilon^{2}
\Bigl(\frac{\pi^{2}}{12} + \frac{1}{2}\Bigr)\Bigr].
\end{eqnarray}
\par The integral $B_{E}(p,\mu)$ can be computed by noting the following identity $B_{E}(p,\mu) = - \frac{p}{2} .  \frac{\partial}{\partial p}\int  \frac{d^{d} q_{1} d^{d} q_{2}}{(q_{1}^{2} + \mu^{2}) 
(q_{2}^{2} + \mu^{2})[(q_{1} + q_{2} + p)^{2} + \mu^{2}]}$.Solving all the loop momentum integrals we obtain
\begin{eqnarray}\label{16}
&&B_{E}(p,\mu) =  \Bigl[\frac{\Gamma(2-\frac{\epsilon}{2})}{2}\Bigr]^{2} \Gamma(-1+\epsilon) \int_{0}^{1} dx [x(1-x)]^{-\frac{\epsilon}{2}}  \nonumber\\
&& \int_{0}^{1} dz z^{\frac{\epsilon}{2} -1} [- \frac{p}{2} .  \frac{\partial}{\partial p}] \Bigl[p^{2}z(1-z) + \mu^{2}\Bigl(1-z \nonumber\\
&&+\frac{z}{x(1-x)}\Bigr)\Bigr]^{1- \epsilon},
\end{eqnarray}
and after straightforward manipulations we get to
\begin{eqnarray}\label{17}
&B_{E}(p,\mu) =  \mu^{-2\epsilon}\frac{p^{2}}{8\epsilon}\Bigl[1-\frac{3}{4}\epsilon - 2\epsilon L_{3E}(p,\mu)\Bigr],
\end{eqnarray}
where $L_{3E}(p,\mu) = \int_{0}^{1}\int_{0}^{1} dx dz (1-z) ln \Bigl[\frac{p^{2}}{\mu^{2}} z(1-z)+ 1- z\nonumber\\ 
+\frac{z}{x(1-x)}\Bigr]$. Substitution of Eqs. (\ref{15}),  and (\ref{17}) into Eq. (\ref{14a}) produces the result
\begin{eqnarray}\label{18}
&I_{3}(p,\mu)|_{PDT} = -\frac{3(\mu^{2})^{1-\epsilon}}{2\epsilon^{2}} \Bigl[1+\frac{1}{2} \epsilon + \epsilon^{2}\Bigl(\frac{\pi^{2}}{12} + 1 \Bigr)\Bigr] \nonumber\\
& -\frac{p^{2}(\mu^{2})^{-\epsilon}}{8\epsilon} \Bigl[1+\frac{1}{4} \epsilon  - 2\epsilon L_{3E}(p,\mu)\Bigr].
\end{eqnarray}
\par Consider the three-loop diagram in Eq. (\ref{4}). Its associated integral is given by
\begin{eqnarray}\label{19}
&I_{5}(p,\mu) = \int \frac{d^{d} q_{1} d^{d} q_{2} d^ {d} q_{3}}{(q_{1}^{2} + \mu^{2})(q_{2}^{2} + \mu^{2})(q_{3}^2 + \mu^ {2})[(q_{1} + q_{2} + p)^{2} + \mu^{2}]}\nonumber\\
& \times \; \frac{1}{[(q_{1} + q_{3} + p)^{2} + \mu^{2}]}.
\end{eqnarray}
\par The appplication of the partial-$p$ operation in this integral yields
\begin{subequations}\label{20}
\begin{eqnarray}
&I_{5}(p) = -\frac{2}{(3d-10)} \Bigl[\mu^{2} (C_{1E}(p,\mu) + 4 C_{2E}(p,\mu)) \nonumber\\
& \qquad + 2 D_{E}(p,\mu)\Bigr],\label{20a}\\
&C_{1E}(p,\mu) = \int \frac{d^{d} q_{1} d^{d} q_{2} d^ {d} q_{3}}{(q_{1}^{2} + \mu^{2})^{2}(q_{2}^{2} + \mu^{2})(q_{3}^2 + \mu^ {2})[(q_{1} + q_{2} + p)^{2} + \mu^{2}]}\nonumber\\
& \times \; \frac{1}{[(q_{1} + q_{3} + p)^{2} + \mu^{2}]},\label{20b}\\
&C_{2E}(p,\mu) = \int \frac{d^{d} q_{1} d^{d} q_{2} d^ {d} q_{3}}{(q_{1}^{2} + \mu^{2})(q_{2}^{2} + \mu^{2})^{2} (q_{3}^2 + \mu^ {2})[(q_{1} + q_{2} + p)^{2} + \mu^{2}]}\nonumber\\
& \times \; \frac{1}{[(q_{1} + q_{3} + p)^{2} + \mu^{2}]},\label{20c}\\
&D_{E}(p,\mu) =  \int  \frac{d^{d} q_{1} d^{d} q_{2} d^{d} q_{3} p.(q_{1} + q_{2} + p)}{(q_{1}^{2} + \mu^{2}) (q_{2}^{2} + \mu^{2})(q_{3}^{2} + \mu^{2})} \nonumber\\
&\qquad \frac{1}{[(q_{1} + q_{2} + p)^{2} + \mu^{2}]^{2}[(q_{1} + q_{3} + p)^{2} + \mu^{2}]}.\label{20d}
\end{eqnarray}
\end{subequations}
\par We employ the homotopy theory arguments by the same token from our discussion for a metric tensor with the Minkowski signature. 
\par We start by computing  $C_{1E}(p,\mu)$ in a more explicit form than that presented in Appendix A.  There are two internal subgraphs corresponding to the integral  $I_{2}(q_{1}+p,\mu)$.  Performing the integration on the loop momenta $q_{2}, q_{3}$, we find
\begin{eqnarray}\label{21}
& C_{1E}(p,\mu) = f(\epsilon) \int_{0}^{1}dx [x(1-x)]^{-\frac{\epsilon}{2}} 
\int_{0}^{1}dy[y(1-y)]^{-\frac{\epsilon}{2}}\nonumber\\
&\int \frac{d^{d} q_{1}}{[q_{1}^{2} + \mu^{2}]^{2}[(q_{1} + p)^{2} + \frac{\mu^{2}}{x(1-x)}]^{\frac{\epsilon}{2}}}\nonumber\\
&\frac{1}{[(q_{1} + p)^{2} + \frac{\mu^{2}}{y(1-y)}]^{\frac{\epsilon}{2}}},
\end{eqnarray}
where $f(\epsilon)= \frac{1}{\epsilon^{2}}\Bigl[1 - \epsilon + \frac{\epsilon^{2}}{2} \psi'(1) + \frac{\epsilon^{2}}{4} \Bigr]$.  By using the identity \cite{Yndu}
\begin{eqnarray}\label{22}
&&\frac{1}{A^{\alpha}B^{\beta}C^{\delta}}= \frac{\Gamma(\alpha+\beta+\gamma)}{\Gamma(\alpha)\Gamma(\beta)\Gamma(\gamma)} \int_{0}^{1}dz z \nonumber\\
&&\int_{0}^{1}\frac{dw u_{1}^{\alpha -1}u_{2}^{\beta -1}u_{3}^{\gamma -1}}{[u_{1}A+u_{2}B+u_{3}C]^{\alpha+\beta+\gamma}},
\end{eqnarray}
where $u_{1}=zw$, $u_{2}=z(1-w)$ and $u_{3}=1-z$, it is easy to show that
\begin{eqnarray}\label{23}
&&C_{1E}(p,\mu)= \frac{\Gamma(2+\epsilon)}{\Gamma^{2}(\frac{\epsilon}{2})} \int_{0}^{1} [x(1-x)]^{-\frac{\epsilon}{2}} dx \int_{0}^{1} [y(1-y)]^{-\frac{\epsilon}{2}} dy\nonumber\\
&&\int_{0}^{1} z^{\frac{\epsilon}{2}+1} (1-z)^{\frac{\epsilon}{2} -1} dz \int_{0}^{1} w 
(1-w)^{\frac{\epsilon}{2} -1} dw \nonumber\\
&&\qquad \times \qquad \int \frac{d^{d}q_{1}}{\Bigl[ F(q_{1}, p, x, y, z, w) \Bigr]^{2+\epsilon}},
\end{eqnarray}
where $F(q_{1}, p, x, y, z, w)=q_{1}^{2} + 2p. q_{1}(1-z w) + p^{2}(1-z w) + \mu^{2}\Bigl[z w +\frac{z(1-w)}{x(1-x)} + \frac{1-z}{y(1-y)} \Bigr] $.
The implementation of the $PDT$ corresponds to computing the loop momentum over $q_{1}$ at the end point singularities of the parametric integrals, namely,  at $z=1$ and $w=1$.  Consequently,  this effectively makes this integral independent of the external momentum.  Expanding the argument of the resulting $\Gamma$ functions originating from the remaining parametric integrals  in terms of $\epsilon =4 -d$ leads to
\begin{eqnarray}\label{24}
&C_{1E}(p,\mu)|_{PDT} = \frac{\mu^{-3\epsilon}}{3\epsilon^{3}}\Bigl[ 1-  \frac{1}{2} \epsilon + 
\frac{3}{4}\epsilon^{2} + \frac{5 \psi'(1)}{4}\epsilon^{2}\Bigr].
\end{eqnarray}
\par Consider $C_{2E}(p, \mu)$.  Integrating over $q_{3}$, we obtain the subdiagram $I_{2}(q_{1}+p)$.  This object can be rewritten in the form
\begin{eqnarray}\label{25}
&& C_{2E}(p,\mu) = \tilde{f}(\epsilon) \int_{0}^{1}dx [x(1-x)]^{-\frac{\epsilon}{2}}\nonumber\\ 
&&\int \frac{d^{d} q_{1} }{[q_{1}^{2} + \mu^{2}][(q_{1} + p)^{2} + \frac{\mu^{2}}{x(1-x)}]^{\frac{\epsilon}{2}}}\nonumber\\
&&\int \frac{d^{d} q_{2}}{[q_{2}^{2} + \mu^{2}]^{2} [(q_{1} + q_{2} + p)^{2} + \mu^{2}]},
\end{eqnarray}
where $\tilde{f}(\epsilon)= \frac{1}{\epsilon}\Bigl[1 - \frac{\epsilon}{2} + \frac{\epsilon^{2}}{4} \psi'(1) \Bigr]$.  Integrating over $q_{2}$, we obtain
\begin{eqnarray}\label{26}
& C_{2E}(p,\mu) = \frac{1}{4}f_{2}(\epsilon) \int_{0}^{1}dx [x(1-x)]^{-\frac{\epsilon}{2}}
\int_{0}^{1} y[y(1-y)]^{-1-\frac{\epsilon}{2}} dy \nonumber\\ 
&\int \frac{d^{d} q_{1} }{[q_{1}^{2} + \mu^{2}][(q_{1} + p)^{2} + \frac{\mu^{2}}{x(1-x)}]^{\frac{\epsilon}{2}}[(q_{1} + p)^{2} + \frac{\mu^{2}}{y(1-y)}]^{1 + \frac{\epsilon}{2}}},
\end{eqnarray}
where $f_{2}(\epsilon) =  \frac{1}{\epsilon}\Bigl[1 - \epsilon + \frac{\epsilon^{2}}{2} \psi'(1) + \frac{\epsilon^{2}}{4} \Bigr]$.  We can now use the identity Eq. (\ref{22}),  and compute the loop integral using $PDT$ , namely at $z=1$ and $w=1$. After simple manipulations, one learns that 
\begin{eqnarray}\label{27}
&C_{2E}(p,\mu)|_{PDT} = - \frac{\mu^{-3\epsilon}}{6\epsilon^{3}}\Bigl[ 1-  \frac{3\epsilon}{2}  + 
\frac{7\epsilon^{2}}{4} + \frac{3 \epsilon^{2} \psi'(1)}{4}\Bigr].
\end{eqnarray}
\par We are left with calculating $D_{E}(p,\mu)$. It is worth emphasizing that this integral can be written as 
\begin{eqnarray}\label{28}
&&2D_{E}(p,\mu) = - \frac{p}{2} . \frac{\partial}{\partial p} \int  \frac{d^{d} q_{1} I_{2}^{2}(q_{1} + p, \mu)}{[q_{1}^{2} + \mu^{2}]}.
\end{eqnarray} 
\par After using the value of the subdiagram outlined in the Appendix A, we can integrate over the loop momentum 
$q_{1}$. The result given just in terms of parametric integrals reads
\begin{eqnarray}\label{29}
&&2D_{E}(p,\mu) = - \frac{p}{2} . \frac{\partial}{\partial p} \Bigl[\frac{-\Gamma(1 -\frac{\epsilon}{2})
\Gamma(\frac{3 \epsilon}{2})}{2 \epsilon^{2} \Gamma(\epsilon)}\Bigr] \int_{0}^{1} dx [x(1-x)]^{-\epsilon}\nonumber\\
&& \int_{0}^{1} dy y^{\epsilon-1}  \Bigl[p^{2} y(1-y) + \mu^{2} \Bigl[ 1-y+ \frac{y}{x(1-x)}\Bigr]\Bigr]^{1 - \frac{3 \epsilon}{2}} . 
\end{eqnarray} 
Performing the derivative, it is not difficult to show that
\begin{eqnarray}\label{30}
&&2D_{E}(p,\mu) = \frac{p^{2} \mu^{-3 \epsilon}}{6 \epsilon^{2}}\Bigl[1 - \epsilon -3 \epsilon L_{3E}(p,\mu) \Bigr],
\end{eqnarray} 
where $L_{3E}(p,\mu)$ was defined after Eq. (\ref{17}).
\par After replacing Eqs. (\ref{24}),(\ref{27}) and (\ref{30}) into Eq. (\ref{20a}),  we find the following expression  
\begin{eqnarray}\label{31}
&I_{5}(p,\mu)|_{PDT} = \frac{\mu^{2-3\epsilon}}{3\epsilon^{3}}\Bigl[ 1-  \epsilon + 
\frac{5}{4}\epsilon^{2} + \frac{ \psi'(1)}{4}\epsilon^{2}\Bigr] - \frac{p^{2} \mu^{-3 \epsilon}}{6 \epsilon^{2}}\Bigl[1 \nonumber\\
&+ \frac{\epsilon}{2} -3 \epsilon L_{3E}(p,\mu) \Bigr].
\end{eqnarray}

\section{Renormalization by minimal subtraction}
\par We proceed to make a separate discussion of the cases with metric tensors with Minkowski and Euclidean signatures.
\subsection{Minkowski signature}
\par The primitively divergent vertex parts $\Gamma^{(2)}$, $\Gamma^{(4)}$ and $\Gamma^{(2,1)}$ suffice to carry out the renormalization of all other vertex parts $\Gamma^{(N,L)}$ that can be renormalized multiplicatively  
($(N,L)\neq (0,0), (0,2)$).
\par From Eqs.(\ref{7}),(\ref{A17}) and (\ref{A28}) we find for $\Gamma^{(2)}$ the following expression:
\begin{eqnarray}\label{32}
&&\Gamma^{(2)} (p,\mu,u_{0}) =  p^{2} - \mu^{2} +  \frac{(N+2)u_{0}^{2} p^{2}}{144 \epsilon}[1 +\frac{\epsilon}{4} \nonumber\\
&&- 2 \epsilon L_{3M}(p,\mu)] - \frac{(N+2)(N+8)u_{0}^{3} p^{2}}{648 \epsilon^{2}}[1+\frac{\epsilon}{2} \nonumber\\
&& - 3 \epsilon L_{3M}(p,\mu)].
\end{eqnarray}
\par The vertex part $\Gamma^{(4)}$ can be written analogously. After using Eq.(\ref{5}) appropriate for Minkowski space in conjunction with Eqs.(\ref{A7}),(\ref{A8}) and (\ref{A10}) we find
\begin{eqnarray}\label{33}
&&\Gamma^{(4)}(p_{i},\mu,\lambda) =  -iu_{0} \mu^{\epsilon} \Bigl[ 1 - \frac{u_{0} (N+8)}{18 \epsilon}\Bigl[ 3 - \frac{3 \epsilon}{2} - \frac{\epsilon}{2}\nonumber\\
&& \times (L_{M}(s, \mu) + L_{M}(t, \mu) + L_{M}(u, \mu))\Bigr] \Bigr]+ u_{0}^{2} \Bigl[\frac{(5N+22)}{54 \epsilon^{2}}(3 \nonumber\\
&& - \frac{3 \epsilon}{2} - \epsilon (L_{M}(s, \mu) + L_{M}(t, \mu) + L_{M}(u, \mu))) \nonumber\\
&& + \frac{(N^{2} +6N+22)}{108 \epsilon^{2}}(3 - 3 \epsilon - \epsilon (L_{M}(s, \mu) + L_{M}(t, \mu) \nonumber\\
&&+ L_{M}(u, \mu)))\Bigr]\Bigr].
\end{eqnarray}
\par  We define the dimensionless bare coupling constant $u_{0}$ and the renormalization functions 
$Z_{\phi}$ and $\bar{Z}_{\phi^{2}} \equiv Z_{\phi} Z_{\phi^{2}}$ as power series in the dimensionless renormalized coupling constant $u$ as 
\begin{subequations}\label{34}
\begin{eqnarray} 
&& u_{0} = u(1+a_{1} u +a_{2} u^{2}), \label{34a} \\
&& Z_{\phi} = 1+b_{2} u^{2} +b_{3} u^{3}, \label{34b} \\
&& \bar{Z}_{\phi^{2}} = 1 + c_{1} u +c_{2} u^{2}, \label{34c} 
\end{eqnarray}
\end{subequations}
\par Multiplicative renormalization states that we can express multiplicatively renormalized vertex parts in terms of the bare vertex parts as  $\Gamma_{R}^{(N,L)}(p_{i};Q_{j},m,u) = Z_{\phi}^{\frac{N}{2}} Z_{\phi^{2}}^{L} \Gamma^{(N,L)}(p_{i};Q_{j},\mu,u_{0})$ ($i=1,...,N$; $j=1,...,L$), where the normalization functions $Z_{\phi}$ and $Z_{\phi^{2}}$ are determined entirely from the finiteness of the renormalized vertex parts obtained from the primitively divergent vertex parts.  
\par Let us apply this prescription first to find the normalization function 
$Z_{\phi}$ and the  dimensionless renormalized coupling constant $u$ at two-loop order. Then,  utilizing the expression $\Gamma_{R}^{(2)}=Z_{\phi} \Gamma^{(2)}(p,\mu,u_{0})$ at two-loops, using $u_{0}^{2} = u^{2}$ and requiring minimal subtraction of dimensional poles, we find 
\begin{equation}\label{35}
b_{2} = -\frac{(N+2)}{144 \epsilon}. 
\end{equation}
After the definition of the finite  renormalized mass in arbitrary loop order by $m^{2} = Z_{\phi} \mu^{2}$, neglecting finite terms we can then write up to two-loop  order $\Gamma_{R}^{(2)}= \Gamma_{R}^{(2)}(p,m,u)$.  We then apply the multiplicative renormalization statement on $\Gamma_{R}^{(4)} = Z_{\phi}^{2} \Gamma^{(4)}$.  First at one loop-order, we obtain $a_{1} = \frac{(N+8)}{6 \epsilon}$.  At two-loops, the simple poles in $\epsilon$ proportional to the coefficients of 
$L_{M}(s,\mu)$,  $L_{M}(t,\mu)$ and $L_{M}(u,\mu)$ vanish simultaneously. In addition, we find  
\begin{equation}\label{36}
a_{2} = \Bigl(\frac{(N+8)}{6 \epsilon}\Bigr)^{2} - \frac{(3N+14)}{24 \epsilon}.  
\end{equation}
\par Now we can determine the coefficient $b_{3}$ at three-loop order. We have to use $u_{0}^{2} = u^{2} + 2 a_{1} u^{3} + O(u^{4})$.  The simple poles, which are the coefficients of $L_{3M}(p,\mu)$, cancel,  and we find 
\begin{equation}\label{37}
b_{3}= - \frac{(N+2)(N+8)}{1296 \epsilon^{2}} 
+ \frac{(N+2)(N+8)}{5184 \epsilon}.
\end{equation}
\par After the definition of the finite  renormalized mass in arbitrary loop order by $m^{2} = Z_{\phi} \mu^{2}$ and neglecting finite terms, we can then write up to three-loop  order $\Gamma_{R}^{(2)}= \Gamma_{R}^{(2)}(p,m,u)$.  A similar argument is valid for $\Gamma_{R}^{(2,1)}$.
\par The argument for $\Gamma_{R}^{(4)}$ is different since it keeps a global factor $\mu^{\epsilon}$.  The way out is to write $\mu^{2}=Z_{\phi}^{-1}m^{2}$ and then set $\epsilon=0$ (or $Z_{\phi}=1$) in the global factor of 
$Z_{\phi}^{\epsilon}$ since this term comes from the tree-level value of $\Gamma_{R}^{(4)}$ which can then be written at two-loops as $\Gamma_{R}^{(4)}=\Gamma_{R}^{(4)}(p_{i},u,m)$. 
\par We close this section by computing $ \bar{Z}_{\phi^{2}}$.  With the values of each diagram indicated in Appendix B, the diagrammatic expansion is 
\begin{subequations}\label{38}
\begin{eqnarray}
&& \Gamma_{R}^{(2,1)}(p_{1},p_{2}; Q,m,u)= [1 +c_{1} u +c_{2} u^{2}][ 1 - C_{1} u_{0} \nonumber\\
&& +(C_{2}^{(1)} + C_{2}^{(2)})u_{0}^{2}], \\
&& C_{1} = \frac{(N+2)}{18 \epsilon}[3(1 - \frac{\epsilon}{2}) - \frac{\epsilon}{2}(L_{M}(s,\mu) + L_{M}(t,\mu) \nonumber\\
&& + L_{M}(u,\mu))],  \\
&& C_{2}^{(1)} = \frac{(N+2)^{2}}{108 \epsilon^{2}}[1 - \epsilon - \epsilon(L_{M}(s,\mu) + L_{M}(t,\mu) \nonumber\\
&& + L_{M}(u,\mu))],\\
&& C_{2}^{(2)} = \frac{(N+2)}{36 \epsilon^{2}}[3(1 - \frac{\epsilon}{2}) - \epsilon(L_{M}(s,\mu) + L_{M}(t,\mu) \nonumber\\
&& + L_{M}(u,\mu))].
\end{eqnarray}
\end{subequations}
\par After the expansion of $u_{0}$ in terms of $u$ combined with the above equations and the required minimal subtraction we first find that all simple poles in $\epsilon$ involving 
$L_{M}(s,\mu), L_{M}(t,\mu)$, and $ L_{M}(u,\mu)$ are canceled. Second,  it is not difficult to show that
\begin{eqnarray}\label{39}
&&\bar{Z}_{\phi^{2}} = 1 + \frac{(N+2)}{6 \epsilon} u + \Bigl[\frac{(N+2)(N+5)}{36 \epsilon^{2}} \nonumber\\
&& - \frac{(N+2)}{24 \epsilon}\Bigr]u^{2}.
\end{eqnarray}
\par Thus we have seen explicitly that the renormalized vertex parts obtained from the primitively divergent bare vertex parts can be written entirely in terms of the external momenta and the renormalized parameters $(u,m)$.  Since an arbitrary vertex part that can be renormalized multiplicatively has an skeleton expansion in terms of the renormalized vertex parts obtained from the primitively divergent bare vertex parts \cite{ZJ},  it is then established 
that $\Gamma_{R}^{(N,L)}=\Gamma_{R}^{(N,L)}(p_{i};Q_{j},m,u)$ at the loop order considered.
\subsection{Euclidean signature}
\par Here, we shall focus directly in the perturbation expansion of the vertex parts $\Gamma^{(2)}(p,\mu,u_{0}), \Gamma^{(4)}(p_{i},\mu,u_{0})$ and 
$\Gamma^{(2)}(p_{1},p_{2}, Q,\mu,u_{0})$ which are given by
\begin{subequations}\label{40}
\begin{eqnarray}
&&\Gamma^{(4)}(p_{i},\mu,u_{0}) =  -u_{0} \mu^{\epsilon} [1  + A_{1} u_{0} + A_{2} u_{0}^{2}], \\
&&\Gamma^{(2)} (p,\mu,u_{0}) =  p^{2} + \mu^{2} + B_{2} u_{0}^{2} + B_{3} u_{0}^{3}, \\
&& \Gamma^{(2,1)}(p_{1},p_{2}; Q,\mu,u_{0})=  1 - C_{1} u_{0} +(C_{2}^{(1)} + C_{2}^{(2)})\nonumber\\
&&\;\times \; u_{0}^{2}, \\
&&A_{1} =  - \frac{(N+8)}{18 \epsilon}\Bigl[ 3 - \frac{3 \epsilon}{2} - \frac{\epsilon}{2}(L_{E}(p_{1} + p_{2}, \mu)\nonumber\\
&& + L_{E}(p_{1} + p_{3}, \mu) + L_{E}(p_{2}+p_{3}, \mu))\Bigr],\\
&&A_{2} = \frac{(5N+22)}{54 \epsilon^{2}}\Bigl(3 - \frac{3 \epsilon}{2} - \epsilon (L_{E}(p_{1} + p_{2}, \mu)\nonumber\\
&& + L_{E}(p_{1} + p_{3}, \mu) + L_{E}(p_{2} + p_{3} \mu))\Bigr) + \frac{(N^{2} +6N+22)}{108 \epsilon^{2}}(3 \nonumber\\
&& - 3 \epsilon - \epsilon (L_{E}(p_{1} + p_{2}, \mu) + L_{E}(p_{1} + p_{3}, \mu)\nonumber\\
&& + L_{E}(p_{2} + p_{3}, \mu))),\\
&&B_{2}= \frac{(N+2) p^{2}}{144 \epsilon}[1 +\frac{\epsilon}{4} - 2 \epsilon L_{3E}(p,\mu)],\\
&&B_{3} = - \frac{(N+2)(N+8)p^{2}}{648 \epsilon^{2}}[1+\frac{\epsilon}{2}  - 3 \epsilon L_{3E}(p,\mu)],\\
&&C_{1} = \frac{(N+2)}{18 \epsilon}[3(1 - \frac{\epsilon}{2}) - \frac{\epsilon}{2}(L_{E}(p_{1} + p_{2},\mu) \nonumber\\
&& + L_{E}(p_{1} + p_{3},\mu) + L_{M}(p_{2} + p_{3},\mu))],  \\
&&C_{2}^{(1)} = \frac{(N+2)^{2}}{108 \epsilon^{2}}[1 - \epsilon - \epsilon(L_{E}(p_{1} + p_{2},\mu) \nonumber\\
&& + L_{E}(p_{1} + p_{3},\mu) + L_{E}(p_{2} + p_{3},\mu))],\\
&&C_{2}^{(2)} = \frac{(N+2)}{36 \epsilon^{2}}[3(1 - \frac{\epsilon}{2}) - \epsilon(L_{E}(p_{1} + p_{2},\mu) \nonumber\\
&& + L_{E}(p_{1} + p_{3},\mu) + L_{E}(p_{2} + p_{3},\mu))].
\end{eqnarray}
\end{subequations}
The requirement of minimal subtraction for the spacetime metric with Euclidean signature is the analog to the case of spacetime metric with Minkowski's signature just discussed: we first compute the coefficient $a_{1} (c_{1})$ using the one-loop diagram from $\Gamma^{(4)}(\Gamma^{(2,1)})$,  then compute $b_{2}$ from the two-loop diagram from $\Gamma^{(2)}$. Next,  compute the coefficients $a_{2},c_{2}$ from $\Gamma^{(4)}, \Gamma^{(2,1)}$ at two-loop order and then the three-loop coefficient $b_{3}$.  The result is 
\begin{subequations}\label{41}
\begin{eqnarray}
&&\bar{Z}_{\phi^{2}}= 1+ \frac{(N+2)}{6 \epsilon} u + \Bigl[\frac{(N+2)(N+5)}{36 \epsilon^{2}} \nonumber\\
&& \quad \quad - \frac{(N+2)}{24 \epsilon}\Bigr] u^{2},\label{41a} \\ 
&& Z_{\phi} = 1 -\frac{(N+2)}{144 \epsilon} u^{2} - \Bigl[\frac{(N+2)(N+8)}{1296 \epsilon^{2}} \nonumber \\
&& \quad \quad - \frac{N+2)(N+8)}{5184 \epsilon}\Bigr] u^{3}. \label{41 b} 
\end{eqnarray}
\end{subequations}
There was a misprint in the sign of the last term in $Z_{\phi} $ in \cite{Leite1} that is now corrected as well.
\par The above results confirm the consistency of the method.  As expected, the renormalization functions in the 
context of Euclidean field theory are {\it the same} as those from the field theory in Minkowski's space,  as we have just shown.  The novelty is that it is the first time that this was shown explicitly with the same set of conventions for both cases,  and using the present minimal subtraction method for a massive scalar field. 
\par We will use the results in this subsection to compute static critical exponents in systems undergoing second-order phase transitions in statistical mechanics shortly. Meanwhile,  next, let us discuss the scale properties of this scalar field theory utilizing a nonperturbative technique.   
\section{The Callan-Symanzik equation from minimal subtraction}
\par The properties of the multiplicatively renormalized vertex functions under a flow in the renormalized parameters, namely the mass and dimensionless coupling constant, are now discussed.  Recall that the renormalized mass and coupling constants in minimal subtraction are no longer defined with normalization conditions at a specified external momentum scale.  Instead, the proposal is to look for a definition of the three-loop renormalized mass solely in terms of three-loop bare mass, in close analogy with the minimal subtraction method to define the renormalized massless theory.  
\par In the latter, the renormalized mass is not defined for a fixed external momentum scale of the two-point vertex part 
$\Gamma_{R}^{(2)}$ as it is in normalization conditions: it is just natural to assign the zero value of the 
renormalized mass (from the massless theory) due to the relation $m^{2}(=0)= Z_{\phi}\mu^{2}(=0)$.  On the other hand, the coupling constant is defined in an arbitrary momentum scale of the renormalized theory 
introduced to make sense on dimensional grounds in $d=4-\epsilon$.  This definition is in contrast with that of the renormalized coupling constant in normalization conditions as the value of the renormalized vertex part 
$\Gamma_{R}^{(4)}$ at a certain set of values of the external momenta scale.
\par These two properties of the renormalized theory in the massless minimal subtraction method suggest an 
analogy to the definition of the same normalization scheme in the massive theory.  
Indeed,  here the renormalized mass $m$ and coupling constant are {\it defined} by $m^{2}=Z_{\phi}\mu^{2}$ and $g=m^{\epsilon} u$ in $d=4-\epsilon$ dimensions, respectively.  We shall restrict our discussion to the Euclidean metric in the next subsections.The case of the Minkowski metric follows the Euclidean one after a Wick rotation in the timelike component of the momentum \cite{W1}.  Nevertheless,  we will explicitly turn our attention to $QFT$,  mainly in the Minkowski space,  at the end of this section to understand the ultraviolet regime at the fixed points.  Eventually, we will also comment on the counterpart of the ultraviolet regime in Euclidean space.
\subsubsection{From $\mu_{0}$ (through $\mu$) to m}
\par The question is how to get to $m$ starting from $\mu_{0}$, the original bare mass parameter in the Lagrangian density. First, let us employ the $PDT$ to outline symbolically $\mu$ as a function of $\mu_{0}$.  To extract the dimensionful dependence of the bare coupling constant, we write $\lambda= \mu_{0}^{-\epsilon} \bar{u}_{0}$. Therefore, we find 
\begin{eqnarray}\label{44}
&& \mu^{2} = \mu_{0}^{2}[1+ \frac{\bar{u}_{0}}{\epsilon}(d_{1}+d_{2} \epsilon) + \frac{\bar{u}_{0}^{2}}{\epsilon^{2}}
 \sum_{i=1}^{2}(d_{1}^{(i)}+d_{2}^{(i)} \epsilon + d_{3}^{(i)} \epsilon^{2}) \nonumber\\
&& + \frac{\bar{u}_{0}^{3}}{\epsilon^{3}} \sum_{i=1}^{5}(d_{1}^{(i)}+d_{2}^{(i)} \epsilon + d_{3}^{(i)} \epsilon^{2} + d_{4}^{(i)} \epsilon^{3})].
\end{eqnarray}
\par When the bare coupling constant is written in terms of $\mu$,  it is given by $\lambda(u_{0},\mu)= \mu^{-\epsilon} u_{0}$, with $u_{0}$ a new value to compensate for the change of scale $\mu_{0} \rightarrow \mu$.  The point is that there is still freedom in the choice of $(\mu_{0}, \bar{u}_{0})$ as divergent quantities: the theory with $\mu$ does not possess any of the tadpole terms.  Therefore, $(\mu_{0}, \bar{u}_{0})$ can be written as power series in $u_{0}$ 
(and inverse power series in $\epsilon$) whose coefficients match exactly those of the theory without tadpoles. The renormalized theory shares the same feature when is written in terms of the renormalized mass $m$.  Writing 
$\lambda=m^{-\epsilon} u_{0}$,  the implicit functional equation obtained from our definition 
$\mu^{2} = Z_{\phi}^{-1}(u,\epsilon) m^{2}$ with $m^{2}$ finite, can be written as $h(u_{0},\epsilon) = f((\mu_{0}/m)^{2} )\times F(\bar{u}_{0}, \epsilon)$.  We do not aim at a rigorous proof of this equation at all loop orders: it suffices to our purposes up to the loop order treated in the present work.
\par The essential point is that, in analogy with the massless theory, once the infinities (poles in $\epsilon$) are removed as shown in the last section,  we can safely replace the bare mass $\mu$ by $m$ in the remainder,  what can always be done provided $\mu$ (or $\mu_{0}$) are nonvanishing.  
\subsubsection{Derivation of the Callan-Symanzik equation}
\par There are two different steps here that provoke a particular deviation from what happens in the massless theory. \par First, treating the perturbative theory without the tadpole contributions,  one can define the three-loop bare mass in terms of the dimensionless bare coupling constant $u_{0}$ as 
$\lambda=\mu^{\epsilon} u_{0}$.  So far,  it looks like as in the massless theory, and whence there is a nonvanishing momentum scale $\kappa$ used to write $\lambda=\kappa^{\epsilon}u_{0}$. When performing the diagrammatic expansion in the massless theory, all the dimensionful polynomials in $\kappa$ cancel out,  and we are left,  after the loop integrations, with the parametric integrals which depend on the dimensionless combination $\frac{p_{i}}{\kappa}$.  This maneuver renders the renormalized theory dependent on $\kappa$. The flow in $\kappa$ originates the renormalization group equations. 
\par After eliminating all polynomial dependence on $\mu$, the second step in the massive theory is to set another mass scale $m$, the renormalized mass, characterizing the renormalized theory. The flow in $m$ will produce the Callan-Symanzik equation in analogy with the renormalization group equation for the massless theory,  as explained above.  Since multiplicative renormalization involves the bare vertex parts, which depend on the bare quantities $\mu,\lambda$ and the flow in $\mu$ (or $m$) comes from the cutoff $\Lambda$ through the dimensionless combination $\frac{\mu}{\Lambda}$ (or $\frac{m}{\Lambda}$) \cite{BLZ,IIM},  it is licit to define the bare coupling constant for scale transformations in the form  $\lambda=m^{\epsilon} u_{0}$ as well as to set in all logarithmic parametric integrals 
$\mu=m$ due to our definition $m^{2}=Z_{\phi}\mu^{2}$.  
\par The extension of the argument of the previous section to arbitrary loop order is straightforward. Consider first the renormalized vertex part $\Gamma_{R}^{(2)}=Z_{\phi} \Gamma^{(2)}(p, \mu, \lambda)$.  As shown in the previous section, after using the "parametric dissociation transform" ($PDT$),  the polynomial dependence of the bare mass $\mu$ is eliminated in the perturbation expansion owing to the unique properties of the perturbative expansion: all polynomial mass dependent terms turn into only one term, namely the renormalized "tree-level" value $m^{2}$ (there are logarithmic corrections in $\mu$ nevertheless).  But we can argue that the parameter $m$ varies due to the definition of a great deal of renormalized theories, each of them constructed from the same bare theory.  Similar arguments can be carried out for $\Gamma_{R}^{(4)}$ and $\Gamma_{R}^{(2,1)}$ such that it is obvious that the renormalized vertex parts are now explicitly dependent on $(m,u)$.  Recall that in arbitrary loop order the renormalized vertex parts which are renormalized multiplicatively are given by $\Gamma_{R}^{(N,L)}=\Gamma_{R}^{(N,L)}(p_{i};Q_{j},m,u)$.
\par The renormalized coupling constant $g$ is dimensionless in four dimensions,  but this is not so in $d=4-\epsilon$ dimensions.  In order to get rid of all the dimensionful flow in the coupling constant, we define the "Gell-Mann-Low" function \cite{NN,V} $\beta_{GL}(g,m)=-\epsilon g + m(\frac{\partial g}{\partial m})_{\lambda,\Lambda}$.  When using this relation, we can express the renormalized theory directly in terms of the dimensionless coupling constant in $d=4-\epsilon$ dimensions, since $\beta_{GL}(g,m)\frac{\partial}{\partial g} = \beta(u)\frac{\partial}{\partial u}$, where $\beta(u)=m(\frac{\partial u}{\partial m})_{\lambda,\Lambda}$.  The cutoff $\Lambda$ is now unnecessary,  and we can write $\lambda=\lambda(u,m)$.  Then one can easily show that $\beta(u)= -\epsilon (\frac{\partial lnu_{0}}{\partial u})^{-1}$.
\par Now consider a flow in $m$ in order to find the following flow equation $\Bigl[m \frac{\partial}{\partial m} + \beta(u) \frac{\partial}{\partial u} -\frac{N}{2} \gamma_{\phi}(u) + L \gamma_{\phi^{2}}\Bigr] \Gamma_{R}^{(N,L)}(p_{i};Q_{j},m,u)= \frac{(2-\gamma_{\phi}(u))m^{2}}{\bar{Z}_{\phi^{2}}}  \Gamma_{R}^{(N,L+1)}(p_{i};Q_{j},0,m,u)$ for the renormalized vertex parts.  We employ the standard notation to denote the renormalization functions by 
$\gamma_{\phi}(u)= \beta(u) \frac{\partial ln Z_{\phi}}{\partial u}$ and $\gamma_{\phi^{2}}(u)= -\beta(u) \frac{\partial ln Z_{\phi^{2}}}{\partial u}$.  Another useful definition is $\bar{\gamma}_{\phi^{2}}(u)= -\beta(u) \frac{\partial ln \bar{Z}_{\phi^{2}}}{\partial u}$. The functions $\gamma_{\phi}(u)$ and $\bar{\gamma}_{\phi^{2}}(u)$ calculated at the nontrivial fixed point will be related to the exponents $\eta$ and $\nu$,  as we shall discuss in a moment.  
\par Please observe that the right-hand side ($RHS$) of the flow equation includes an extra factor ${\bar{Z}_{\phi^{2}}}$ in the denominator since we did not use normalization conditions to fix this term.  In order to achieve "covariance under the renormalization scheme" for the solution of the flow equation when either using normalization conditions or minimal subtraction as proposed herein, we set the tree-level value ${\bar{Z}_{\phi^{2}}}=1$ in the $RHS$ in order to obtain the Callan-Symanzik $(CS)$ equation
\begin{eqnarray}\label{45}
&\Bigl[m \frac{\partial}{\partial m} + \beta(u) \frac{\partial}{\partial u} -\frac{N}{2} \gamma_{\phi}(u) + L \gamma_{\phi^{2}}\Bigr] \Gamma_{R}^{(N,L)}(p_{i};Q_{j},m,u)= \nonumber\\
&(2-\gamma_{\phi}(u))m^{2}  \Gamma_{R}^{(N,L+1)}(p_{i};Q_{j},0,m,u).
\end{eqnarray} 
\par The scaling region of the solutions to this equation can now be stated. It can be understood from the fact that (the insertion at zero momentum represented by) the vertex parts  $\Gamma_{R}^{(N,L+1)}(p_{i};Q_{j},0,m,u)$ have an extra propagator in comparison 
with $\Gamma_{R}^{(N,L)}(p_{i};Q_{j},m,u)$ so it goes to zero more rapidly than the latter in the ultraviolet ($UV$)  limit $\frac{|p|}{m} \rightarrow \infty $.  In the UV regime, the $RHS$ of this equation is neglected in comparison with the left-hand side ($LHS$) \cite{W1}, therefore validating the scaling hypothesis \cite{Amit}.  
\par Note that without the $PDT$ the two-point vertex function in the unconventional minimal subtraction 
presented in  \cite{CarLei} for the Euclidean case does not satisfy the $CS$ equation above.  After renormalizing $\Gamma_{R}^{(2,1)}$ and 
$\Gamma_{R}^{(4)}$ in two-loops, there was an awkward extra subtraction prescribed in that method to remove an extra simple pole in $\epsilon$ proportional to the mass at three-loop order for $\Gamma_{R}^{(2)}(k, u, m)$. To be specific this unconventional vertex part reads:
\begin{eqnarray}\label{46}
&& \tilde{\Gamma}_{R}^{(2)}(k, u, m) = \Gamma_{R}^{(2)}(k, u, m) + m^{2}u^{3}\frac{(N+2)(N+8)\tilde{I}(k)}{108\epsilon} ,\nonumber\\
&& \tilde{I}(k) = \int_{0}^{1} dx \int_{0}^{1}dy lny 
\frac{d}{dy}\Bigl((1-y) \times \;\nonumber\\
&& ln\Bigl[\frac{y(1-y)\frac{k^{2}}{\mu^{2}} + 1-y 
+ \frac{y}{x(1-x)}}{1-y 
+ \frac{y}{x(1-x)}} \Bigr]\Bigr).
\end{eqnarray} 
\par Differently from the method just presented herein utilizing the $PDT$, the former method possesses the extra logarithmic parametric integral above.  In this particular case, we could have set $\mu^{2}=m^{2}$ inside the residual logarithm integral, but since the extra term has a coefficient proportional to $m^{2} u^{3}$ a simplified version of the flow equation on the vertex  part $\tilde{\Gamma}_{R}^{(2)}(p, u, m)$ would produce instead the following equation 
\begin{eqnarray}\label{47}
&\Bigl[m \frac{\partial}{\partial m} + \beta(u) \frac{\partial}{\partial u} -\frac{N}{2} \gamma_{\phi}(u) + L \gamma_{\phi^{2}}\Bigr] \Gamma_{R}^{(2)}(p,m,u)= \nonumber\\
&(2-\gamma_{\phi}(u))m^{2}  \Gamma_{R}^{(2,1)}(p;0,m,u) + \Bigl[(2m^{2} u^{3} + 3\beta(u) \nonumber\\
& u^{2} m^{2})\tilde{I}(p) + m^{2}u^{3} \frac{\partial \tilde{I}(p)}{\partial m^{2}}\Bigr]\frac{(N+2)(N+8)}{108\epsilon} .
\end{eqnarray} 
\par Therefore, although we can still apply the ultraviolet regime to get rid of the first term in the $RHS$ of the last equation, the presence of the pole in $\epsilon$ (which is dominant even in the ultraviolet regime) prevents 
$\Gamma_{R}^{(2)}$ from having a scaling regime. This undesirable state of affairs signals an inconsistency of the unconventional minimal subtraction method in three-loop order.  Note that the present $PDT$ method does not suffer from this pathology.
\par The $CS$ equation is valid at all orders in perturbation theory.  In a generic loop order, the argument outlined here up to three-loops explicitly for the vertex part $\Gamma_{R}^{(2)}$ can be extended utilizing similar arguments.  We start with the bare mass $\mu_{0}$, and define the bare mass at $(L+1)th$-loop order as the value of the bare vertex 
$\Gamma^{(2)}$ at zero external momentum.  The replacement $\mu_{0}^{2} = \mu^{2} + ...$ eliminates all tadpole diagrams up to $(L+1)th$-loop order. This vertex part can be dealt with by using $PDT$ such that all the mass-dependent terms reduce only to the tree-level value.  With the definition of $(L+1)th$-loop order renormalized mass,  this vertex part at $(L+1)th$-loop order and the other two primitively divergent vertex parts at $Lth$-loop order can all be renormalized by minimal subtraction as described. Consequently, all the multiplicatively renormalized vertex parts can be renormalized by minimal subtraction and satisfy the above  $CS$ equation.
\subsubsection{The analysis of the ultraviolet region using the Callan-Symanzik equation}
\par  Now we would like to  investigate the ultraviolet region of the theory by applying the Callan-Symanzik equation just obtained in a manifest mass-independent way.  We will consider metric tensors with Minkowski (corresponding to a true $QFT$) and  Euclidean (statistical mechanics) signatures.  This strategy will help to see whether the pictures correspond to actual ultraviolet fixed points.
\par Typical scale-invariant phenomena occur in the ultrarelativistic scattering processes in elementary particle physics ($QFT$ in the ultraviolet region in $d=4$).  For instance, in the deep inelastic scattering of electrons and protons,  Bjorken scaling for the analog of the vertex part $\Gamma_{R}^{(4)}$ herein takes place for high energies \cite{Bjorken}.  The intermediate particle is a photon and its momentum is off-shell and spacelike.  Moreover,  Higgs bosons with spacelike momenta can be exchanged in a different kinematical (higher energy) regime from this example of deep inelastic scattering\cite{SMVV}.
\par \par Previous $QFT$ studies for $d=4$ identified what would be a high energy regime.  Indeed, the incoming and outgoing fields in the four-point function were set at off-shell and spacelike external momenta to characterize inclusive annihilation processes in $\phi^{4}$ field theory as the analog of deep inelastic scattering \cite{Mue}.  However,  there is was no physical reason for this choice of external (and internal) momenta in that work. (The Callan-Symanzik equation was used in that work to renormalize the scattering amplitude, but the analysis was not carried out at the fixed point.) It is probably fair to say that a field configuration with spacelike momenta characterizes an intermediate state from a deep inelastic scattering process.   
\par In $d=4-\epsilon$  dimensions, (with $\epsilon \neq 0$) the Callan-Symanzik is an invaluable tool to obtain information about anomalous dimensions of certain operators \cite{Mitter} utilizing the operator product expansion ($OPE$) \cite{W3} at the fixed point.  In particular, the deep inelastic scattering region within this analysis was restricted to Euclidean momentum.  
\par Consider the fixed points, defined by the equation $\beta(u_{fp})=0$.  The trivial one corresponds to $u_{fp}=0$ which yields $\gamma_{\phi}(u_{fp})=0$ ($d\neq 4$).  With the simplest application in mind,  let us use the $CS$ equation in the elementary building block in perturbation theory: the propagator (or its inverse).  For the sake of simplicity, we can pick out only the leading order term ($O(u^{0})$) for $\Gamma_{R}^{(2)}$. 
Let us replace this value at the $CS$ equation for $\Gamma_{R}^{(2)}= p^{2} - m^{2}$.  Recalling that, at the fixed points, the $RHS$ of the $CS$ equation can be neglected, we find $m=0$.  So if we start with a massive $\phi^{4}$ theory,  $u=0$ is {\it not} an ultraviolet fixed point since it is forbidden by the $CS$ equation.  This observation agrees with a well-known fact: ultraviolet asymptotic free theories only exist for nonAbelian massless field theories \cite{GW1,P1} in $d=4$ (see also \cite{CG,GW2,P2}). 
\par The same conclusion is valid for the Euclidean case: $u=0$ cannot be an ultraviolet fixed point of a massive theory since the $CS$ equation implies $m=0$, contradicting the initial hypothesis ($m \neq 0$).  In comparison with the massless theory,  an infrared Gaussian fixed point for $d<4$ exists but is not stable. The criterion for instability in the infrared regime transliterates itself on the impossibility of the trivial fixed point being a true ultraviolet fixed point in the massive theory.  
\par Now consider the nontrivial fixed point $\beta(u_{\infty})=0$.  Since we proved that the Minkowski and the Euclidean versions of the metric tensor have the same normalization functions $Z_{\phi}, Z_{\phi^{2}}$ and $u_{0}(u)$,  they define the same set of the Wilson functions $\beta(u)$, $\gamma_{\phi}(u)$ ($\bar{\gamma}_{\phi^{2}}$ will be discussed in next section).  The Wilson function $\beta(u) = -\epsilon \left(\frac{\partial lnu_{0}}{\partial u}\right)^{-1}$  can be expressed in terms of the renormalization coefficients up to two-loop order as $\beta(u) = -\epsilon u[1- a_{1} u + 2(a_{1}^{2} - a_{2})u^{2}]$.  By replacing the coefficients in this expression, we find
\begin{eqnarray}\label{48}
&& \beta(u)=  = u \Bigl[-\epsilon + \frac{(N+8)}{6} u \nonumber\\
&& \qquad \qquad - \frac{(3N+14)}{12} u^{2}\Bigr].
\end{eqnarray}
\par The nontrivial UV fixed point is defined by  
$\beta(u_{\infty})=0$, which implies that $u_{\infty} = \frac{6 \epsilon}{(N+8)}\Bigl[1 + \frac{3(3N+14)}{(N+8)^{2}} \epsilon \Bigr]$.  For the time being, we just need the function $\gamma_{\phi} (u)$.  In terms of the coefficients previously found in the minimal subtraction algorithm above described, this functions have the following expansion
$\gamma_{\phi} (u)= - \epsilon u[2b_{2} u +(3b_{3} - 2b_{2} a_{1})u^{2}]$. At the fixed point it corresponds to the anomalous dimension of the field, namely $\gamma_{\phi} (u_{\infty})=\eta$. As emphasized before, the anomalous dimension is the same for Minkowski as well as Euclidean massive $\phi^{4}$ field theory.  It is given by
\begin{equation}\label{49}
\eta= \frac{(N+2)\epsilon^{2}}{2(N+8)^{2}}\Bigl[1+\Bigl(\frac{6(3N+14)}{(N+8)^{2}} -\frac{1}{4}\Bigr)\epsilon \Bigr].
\end{equation}
\par Consider Minkowski signature. The $CS$ equation for $\Gamma_{R}^{(2)}$ at this fixed point can be written as
\begin{eqnarray}\label{50}
&\Bigl[m \frac{\partial}{\partial m} -\eta] \Gamma_{R}^{(2)}(p,m,u)= 0.
\end{eqnarray} 
\par  The above $CS$ equation is satisfied provided that ($Q^{2} = -p^{2}$ for $p^{2}<0$) 
\begin{equation}\label{51}
\frac{Q^{2}}{m^{2}} = \frac{2}{\eta} -1 \approx 4 \frac{(N+8)^{2}}{(N+2)\epsilon^{2}}.
\end{equation}
\par Then, the $CS$ equation yields automatically a spacelike momentum at the fixed point.  Since the only 
scale-invariant phenomenon with spacelike momentum corresponds to deep inelastic scattering, we conclude that the fixed point is due to the ultraviolet regime.  
\par For the Euclidean version,  one just has to make the replacement $-p^{2} \rightarrow p^{2}$ and there is 
a regime indicating the "high energy" limit in the context of critical phenomena with large (but finite) correlation length: 
it is characterized by space distances $x$ much smaller than the correlation length $\xi$, namely 
$\frac{p}{m} \approx \frac{\xi}{x} \rightarrow \infty$.
Thus the Minkowski and Euclidean versions have similar "high energy" behavior. 
\par In Minkowski spacetime,  last equation is consistent with the kinematical limit of the massive $\phi^{4}$ field theory in the region of high energy behavior of scattering amplitudes for the scalar theory using $OPE$ techniques \cite{C1,Mitter} which requires asymptotic states with large spacelike momentum \cite{W1} for small 
$\epsilon (\neq 0)$.  Second,  a parallel claim is valid in the case of spacetime metric with Euclidean signature.  The high energy behaviors of the Minkowski and Euclidean signature cases are analogous in all respects as long as the ultraviolet behavior is concerned (see also \cite{Mue}).     
\par Thus, the nontrivial fixed point supports the idea of a true ultraviolet fixed point for the massive $\phi^{4}$ $QFT$ (Minkowski or Euclidean) in $d=4-\epsilon$.  This observation starkly contrasts with the situation in four dimensions with some indication for triviality of massive $\phi^{4}$ theory \cite{Wilson2,Call}.  One should be careful about any definitive claim for triviality: observe that setting $\epsilon=0$ 
$(d=4)$ in the $\epsilon$-expansion does not reduce naturally to the four-dimensional case, as the perturbation expansion is no longer valid, even in the massless case \cite{ARR}.
\par The ultraviolet,  nontrivial fixed point is not attractive: the theory is scale-invariant if the renormalized coupling constant is fixed precisely at this value \cite{BLZ,Mitter}.  A curious aspect is that the Wilson functions and the nontrivial fixed points are {\it the same} in the massless and the massive case.  This aspect should not be surprising: minimal subtraction only requires the existence of a nonvanishing scale for the flow of the renormalized theory.  In the massless theory it is a momentum scale whereas in the massive case is the mass scale.  Both of them have momentum units: the equality between them is just a consequence of the scale-independence of minimal subtraction.  Ultraviolet divergences are eliminated in the same way in both cases.  However, the infrared divergences which are physical and provides the basis for the flow of the massless theory from the trivial (Gaussian) to the nontrivial fixed points are not present in the massive theory. Therefore, although having both fixed points only one is rigorously ultraviolet: there is no flow between them under scaling transformations. 
\par The massless theory has the same behavior in the infrared nontrivial fixed point as the massive 
theory in the ultraviolet nontrivial fixed point because the mass is negligible in the latter.  One  should not be confused with the misleading claim that  $u_{\infty}$ is the infrared fixed point in the massive theory.  Both fixed points have the same value,  but were obtained from quite a different scaling regime as proved in the present work.  
\section{The computation of critical exponents $\eta$ and $\nu$}
\par We now apply the method in Euclidean space to compute the static critical exponents $\eta$ and $\nu$ from the diagrammatic expansion in the case of a scalar field with an $O(N)$ internal symmetry.  
\par We will be a bit redundant in this section just to make sure to the reader that this corresponds to the application of the aforementioned formalism in the statistical mechanics of critical phenomena. Here, the renormalized mass can be taken in terms of the reduced temperature $m^{2} = t$ with $t=\frac{|T-T_{c}|}{T_{c}}$.  This means that we are going to compute the critical exponents away from the critical temperature.  Note that the regime
$|\frac{p}{m}| \rightarrow \infty$ is equivalent to computing the correlation functions at spatial distances much smaller than the correlation length, which is finite away from the critical temperature. These exponents are examples of universal quantities in the universality class $(N,d)$ of critical systems undergoing a second-order phase transition.
\par Just as before (in the Minkowski case) we can write up to two-loop order $\beta(u) = -\epsilon u[1- a_{1} u + 2(a_{1}^{2} - a_{2})u^{2}]$ and obtain 
\begin{eqnarray}\label{52}
&& \beta(u)=  = u \Bigl[-\epsilon + \frac{(N+8)}{6} u \nonumber\\
&& \qquad \qquad - \frac{(3N+14)}{12} u^{2}\Bigr].
\end{eqnarray}
\par First, there is a trivial fixed point $u=0$, the Gaussian fixed point.  The fixed point that yields the physical quantities of interest like anomalous dimension, etc., is the nontrivial one.  The nontrivial (repulsive) UV fixed point is defined by  
$\beta(u_{\infty}=0$), which implies that $u_{\infty} = \frac{6 \epsilon}{(N+8)}\Bigl[1 + \frac{3(3N+14)}{(N+8)^{2}} \epsilon \Bigr]$. 
\par The aforementioned critical exponents are directly related to the renormalization group functions $\gamma_{\phi}(u)$ and $\bar{\gamma}_{\phi^{2}}(u)$ at the ultraviolet fixed point $u_{\infty}$.  In terms of the coefficients previously found in the minimal subtraction algorithm above described, these functions have the following expansions
$\gamma_{\phi} (u)= - \epsilon u[2b_{2} u +(3b_{3} - 2b_{2} a_{1})u^{2}]$ and $\bar{\gamma}_{\phi^{2}} (u)= \epsilon u[c_{1} + (2c_{2} - c_{1}^{2} - a_{1}c_{1})u]$. From the solution to the coefficients we find
\begin{subequations}
\begin{eqnarray}\label{53}
&& \gamma_{\phi} (u)= \frac{(N+2)}{72}u^{2} -\frac{(N+2)(N+8)u^{3}}{1728} ,\label{53a}\\
&& \bar{\gamma}_{\phi^{2}} (u)=\frac{(N+2)}{6}u\Bigl[1-\frac{u}{2} \Bigr].\label{53b}
\end{eqnarray}
\end{subequations}
\par When computed at the fixed point,  the first one yields the exponent 
$\eta(\epsilon)= \gamma_{\phi}(u_{\infty})$ up to three-loop order, namely
\begin{equation}\label{54}
\eta= \frac{(N+2)\epsilon^{2}}{2(N+8)^{2}}\Bigl[1+\Bigl(\frac{6(3N+14)}{(N+8)^{2}} -\frac{1}{4}\Bigr)\epsilon \Bigr].
\end{equation}
\par Furthermore, the function $\bar{\gamma}_{\phi^{2}}(u)$ calculated at the fixed point permits to determine  the correlation length exponent $\nu(\epsilon)$ by using the identity 
$\nu^{-1}= 2 - \bar{\gamma}_{\phi^{2}} (u_{\infty}) - \eta$.  It is given by the expression
\begin{equation}\label{55}
\nu= \frac{1}{2} + \frac{(N+2)}{4(N+8)} \epsilon + \frac{(N+2)(N^{2} + 23N + 60)}{8(N+8)^{3}} \epsilon^{2}. 
\end{equation}
\par Note that here, due to the condition $|\frac{p}{m}| \rightarrow \infty$,  the mass can be neglected: the asymptotic ultraviolet limit is the same as the infrared limit of the massless theory.  
\par In normalization conditions, the renormalization functions are all different when one compares massless and massive theories: even the nontrivial fixed points are different.  For instance,  $\gamma_{\phi} (u)$ and 
$\bar{\gamma}_{\phi^{2}}(u)$ are identical solely at the nontrivial fixed point(s) in the massless and massive renormalized theories.  This is the well-known manifestation of universality.

\par 
\section{Conclusions}
\par The method presented provides a detailed discussion of that introduced in ref.  \cite{Leite1} concerning the case of an Euclidean $\lambda\phi^{4}$ field theory. In addition, we included a parallel treatment of a similar quantum field theory with the Minkowski signature for the spacetime metric tensor. This treatment completes the unified picture of the most straightforward massive minimal subtraction scheme in the statistical mechanics of static critical phenomena and relativistic applications occurring, for instance, in particle physics.  
\par From the $QFT$ viewpoint of a massive $\phi^{4}$ theory,  we have just proved that the application of the $CS$ equation to the renormalized two-point function at the nontrivial fixed point implies that the momentum of the associated propagator is spacelike for $d \neq 4$ in that regime.  Its existence can only be guaranteed if the scalar particle associated with the field is in an internal state, like a Higgs boson in some kinematical regions occurring in deep inelastic electron-nucleon scattering,  as described in \cite{SMVV}.   Remarkably, the Born term in perturbation theory,  as employed herein,  is sufficient to predict the existence of the kinematical configuration analogous to deep inelastic scattering at the nontrivial fixed point.  It is then consistent with a true ultraviolet (fixed point) behavior.  From the statistical mechanics viewpoint, the nontrivial fixed point occurs in the ultraviolet regime characterized by space distances much smaller than the correlation length.  The method has potential applications in both scenarios mentioned above.
\par As it turns out, the technique described can efficiently be used to perform renormalized perturbative computations in the statistical mechanics of finite-size systems in a parallel plate geometry of infinite extent along $(d-1)$ directions and finite size $L$ perpendicular to that hyperplane.  A recent proposal could address critical properties of finite-size systems in momentum space when the order parameter satisfies diverse boundary conditions using normalization conditions.  The massless minimal subtraction version treated only the cases associated with the field's periodic and antiperiodic boundary conditions at the limiting surfaces at, say, $z=0,L$ \cite{BL}.  It would be trivial to extend the present method to these boundary conditions in the massive case.
\par Less obvious is how to extend the techniques presented herein to Neuman and Dirichlet boundary conditions \cite{SBL}. Nevertheless, the techniques presented could be employed to tackle these problems from the perspective of minimal subtraction.  
\par We would like to acknowledge partial support from CAPES (Brazilian agency) through the PROEX Program grant number 0041/2022. 
\appendix
\section{Feynman diagrams in Minkowski space}

\par We are going to write explicit expressions for the diagrams in Minkowski space in order to make the discussion self-contained.  Observe that any prefactor included in the definition of the integration measure in momentum space is equally valid in the perturbation expansion in powers of the coupling constant.  Since $\Gamma^{(4)}$ starts with 
$-i\lambda$,  the prefactor can easily be absorbed in a redefinition of $\lambda$.
\par Now, the origin of our present convention in adopting a suitable prefactor for each loop integral can be 
understood in terms of the well-known result
\begin{equation}\label{A1}
\int d^{d}x exp[-Ax^{2}] = \pi^{\frac{d}{2}} A^{-\frac{d}{2}}.
\end{equation} 
By a redefinition of the measure of the $d$-dimensional integral $d^{d}x \equiv  \frac{d^{d}x}{(2 \pi)^{d} (AF)}$, where 
(Amit's factor) $AF= [2^{d-1} \pi^{\frac{d}{2}} \Gamma(\frac{d}{2})]^{-1}$, the previous integral results in the expression
\begin{equation}\label{A2}
\int d^{d}x exp[-Ax^{2}] = \frac{1}{2} \Gamma(\frac{d}{2}) A^{-\frac{d}{2}}.
\end{equation}
\par The effect of the measure redefinition in momentum space is achieved by making the replacement $\pi^{\frac{d}{2}} \rightarrow \frac{1}{2} \Gamma(\frac{d}{2})$ \cite{Ry}.  
\par Consider the integral $I_{\alpha}(k, \mu)$ defined by
 \begin{eqnarray}\label{A3}
& I_{\alpha}(k, \mu) = \int \frac{d^{d} q}{[q^{2}+2k.q- \mu^{2}]^{\alpha}}  .
\end{eqnarray}
Performing this integral in the usual way first over the timelike momentum $q_{0}$ and then over the spacelike momentum $\vec{q}$,  it is easy to show that  
\begin{eqnarray}\label{A4}
& I_{\alpha}(k, \mu) = i (-1)^{\alpha} \frac{\Gamma(\frac{d}{2}) \Gamma(\alpha - \frac{d}{2})}{2 \Gamma(\alpha)} (\mu^{2} + k^{2})^{\frac{d}{2} -\alpha},
\end{eqnarray}
which can also be rewritten as
\begin{eqnarray}\label{A5}
& I_{\alpha}(k, \mu) = i (-1)^{\frac{d}{2}} \frac{\Gamma(\frac{d}{2}) \Gamma(\alpha - \frac{d}{2})}{2 \Gamma(\alpha)} (-\mu^{2} - k^{2})^{\frac{d}{2} -\alpha}.
\end{eqnarray}
In general,  when expanding in $d=4-\epsilon$,  we set $\epsilon=0$ in the exponent of the overall factor $(-1)$ in the above formulae.
\par Taking a slight deviation from the notation employed so far, let us classify the diagrams into integrals, which count the number of propagators they contain.  Consider first $\Gamma^{(4)}$.  The one-loop diagram is written as a prefactor times $I_{2}$.  By writing $\lambda = \mu^{\epsilon}u_{0}$ ($u_{0}$ is the dimensionless bare coupling constant), we have  
\begin{subequations}  
\begin{eqnarray}\label{A6}
& \parbox{10mm}{\includegraphics[scale=0.9]{fig10.eps}}\quad = u_{0}^{2} \mu^{2\epsilon} \frac{(N+8)}{18} I_{2},\\
& I_{2} = \int \frac{d^{d} q}{[q^{2}- \mu^{2}][(q+P)^{2} - \mu^{2}]}.
\end{eqnarray}
\end{subequations}
\par The integral can be performed using Feynman parameters along with the formulas given above.  The solution in terms of $\epsilon= 4 - d$ is given by 
\begin{subequations}
\begin{eqnarray}\label{A7}
& \parbox{10mm}{\includegraphics[scale=0.9]{fig10.eps}}\quad = i u_{0}^{2} \mu^{\epsilon} \frac{(N+8)}{18\epsilon} \Bigl[ 1 - \frac{\epsilon}{2} - \frac{\epsilon}{2} L_{M}(P, \mu)\Bigr],\label{A7a}\\
& L_{M}(P,\mu) = \int_{0}^{1} dx ln\Bigl[\frac{P^{2}}{\mu^{2}} x(1-x) - 1 \Bigr],\label{A7b}
\end{eqnarray}
\end{subequations}
where $ L_{M}(P,\mu)$ $(\equiv L_{M})$ is finite.  Here, $P$ corresponds to the combinations of external momenta $p_{1} + p_{2}$, $p_{2} + p_{3}$ and $p_{1} + p_{3}$, whose squares are  associated with the Mandelstam variables $s$, $t$, and $u$ in Minkowski space, respectively. 
\par The trivial two-loop graph of the four-point vertex corresponds to the expression $ i u_{0}^{3} \mu^{3\epsilon}\frac{(N^{2}+6N+20)}{108}I_{2}^{2}$, which can be written as 
\begin{eqnarray}\label{A8}
& \parbox{10mm}{\includegraphics[scale=0.9]{fig11.eps}}\quad = -i u_{0}^{3} \mu^{\epsilon}\frac{(N^{2}+6N+20)}{108\epsilon^{2}}  \Bigl[1 -\epsilon -\epsilon L_{M}\Bigr].
\end{eqnarray}
\par The nontrivial diagram of the two-loop contribution for the four-point vertex part is  
$ i u_{0}^{3} \mu^{3 \epsilon}\frac{5N+22}{54} I_{4}$, which reads
\begin{eqnarray}\label{A9}
&& \parbox{10mm}{\includegraphics[scale=0.8]{fig12.eps}}\quad = i u_{0}^{3} \mu^{3 \epsilon} \frac{(5N+22)}{54}  \int \frac{d^{d} q_{1} d^{d} q_{2}}
{[q_{1}^{2} - \mu^{2}][q_{2}^{2} - \mu^{2}]}\nonumber\\
&& \times \frac{1}{[(P-q_{1})^{2} - \mu^{2}][(q_{1}- q_{2} + p_{3})^{2} - \mu^{2}]}.
\end{eqnarray} 
Its $\epsilon$-expansion is given by
\begin{eqnarray}\label{A10}
& \parbox{10mm}{\includegraphics[scale=0.8]{fig12.eps}} = \frac{-i u_{0}^{3} \mu^{\epsilon} (5N+22)}{108\epsilon^{2}} [ 1 - \frac {1}{2}\epsilon - \epsilon L_{M}(P,\mu)] .
\end{eqnarray} 
\par Next, we express the diagrams of $\Gamma^{(2,1)}(p_{1},p_{2};Q,\mu,u_{0})$ in terms of the integrals appearing in the several diagrams of $\Gamma^{(4)}$ (taking into account the summation of contributions from the $s,t$ and $u$ channels as well).  The last two-loop graph shows that the singular part of $I_{4}(p_{i}, \mu)$ depends only on the combinations $s,t$, and $u$.  Loosely speaking, when considering only the singular part of this integral, we can write 
$I_{4}(p_{i}, \mu)= I_{4}(P, \mu)$.  Consider the contribution of the $s$ channel for each diagram from $\Gamma^{(2,1)}(p_{1},p_{2};Q,\mu,u_{0})$. They can be written as
\begin{eqnarray}\label{A11}
&\parbox{10mm}{\includegraphics[scale=0.8]{fig14.eps}}= i u_{0} \mu^{\epsilon} \frac{(N+2)}{18}I_{2} (s,\mu)\nonumber\\
& \parbox{10mm}{\includegraphics[scale=0.8]{fig16.eps}} = - u_{0}^{2} \mu^{2\epsilon} \frac{(N+2)^{2}}{108}
I_{2}^{2} (s,\mu) \nonumber\\
&\parbox{10mm}{\includegraphics[scale=0.8]{fig17.eps}} =  - u_{0}^{2} \mu^{2\epsilon} \frac{(N+2)}{72}
I_{4}^{2} (s,\mu).
\end{eqnarray}  
This expression will suffice for our purposes and we let the reader to work out each term with the values already computed above.  
\par Now we focus our attention on the solution of the two- and three-loop contributions to the two-point vertex part without including any tadpole contributions, as explained in the main text.  The two-loop diagram is represented by the product $\frac{i u_{0}^{2} \mu^{2 \epsilon}(N+2)}{18}I_{3}$,  where $I_{3}$ for Minkowski space reads
\begin{eqnarray}\label{A12}
& I_{3}= 
\int \frac{d^{d} q_{1} d^{d} q_{2}}{[q_{1}^{2} - \mu^{2}][q_{2}^{2} - \mu^{2}]} \nonumber\\
& \;\;\; \times \frac{1}{[(q_{1} + q_{2} + p)^{2}) - \mu^{2}]}.
\end{eqnarray}
After using the partial-$p$ operation, as explained in Section III,  the integral can be written in the form
\begin{subequations}\label{A13}
 \begin{eqnarray}
& I_{3} = \frac{1}{(d-3)} \Bigl[3 \mu^ {2} A_{M}(p,\mu)- B_{M}(p,\mu) \Bigr], \label{A13a}\\
&A_{M}(p,\mu) = \int  \frac{d^{d} q_{1} d^{d} q_{2}}{[q_{1}^{2} - \mu^{2}]^{2}[q_{2}^{2} - \mu^{2}][(q_{1} + q_{2} + p)^{2} - \mu^{2}]}, \label{A13b}\\
&B_{M}(p,\mu) = \int  \frac{d^{d} q_{1} d^{d}q_{2} p. (q_{1} +q_{2} +p)}{[q_{1}^{2} - \mu^{2}]
[q_{2}^{2} - \mu^{2})[(q_{1} + q_{2} + p)^{2} - \mu^{2}]^{2}}.  \label{A13c}
\end{eqnarray}
\end{subequations}
The first integral can be written as 
$A_{M}(p,\mu) = \int  \frac{d^{d} q_{1}}{[q_{1}^{2} - \mu^{2}]^{2}} I_{2}(q_{1} +p)$.  By employing the solution for $I_{2}$, along with the introduction of another Feynman parameter to produce a single denominator,  this joint maneuver leads to
\begin{eqnarray}\label{A14}
&& A_{M}(p,\mu) = i \frac{\Gamma(2-\frac{\epsilon}{2})\Gamma(2+\frac{\epsilon}{2}) }{2} \int_{0}^{1}dx [x(1-x)]^{-\frac{\epsilon}{2}} \int_{0}^{1} dy \nonumber\\ 
&& \;\; \times \int \frac{y^{\frac{\epsilon}{2}-1} (1-y)d^{d} q_{1}}{[q_{1}^{2} + 2 q_{1}.p y + p^{2}y - \mu^{2}(1-y + \frac{y}{x(1-x)})]^{2}}.
\end{eqnarray}
As explained in the main text, we compute the momentum integral at $y=0$ ($PDT$) which has the virtue of eliminating all the dependence in the external momentum $p$.  Solving the parametric integrals after this step and expanding in $d=4-\epsilon$, we find ($A_{M}(p,\mu)_{PDT} \equiv A_{M}(\mu)$)
\begin{equation}\label{A15}
A_{M}(\mu)= -\frac{\mu^{-2\epsilon}}{2 \epsilon^{2}} \Bigl[1 - \frac{\epsilon}{2} + \epsilon^{2}(\frac{1}{2} + \frac{\pi^{2}}{12})\Bigr].
\end{equation} 
\par The integral $B_{M}(p,\mu)$ can be easily computed and we find 
\begin{subequations}
\begin{eqnarray}\label{A16}
&& B_{M}(p,\mu) = - \frac{p^{2} \mu^{-2\epsilon}}{8 \epsilon} \Bigl[ 1 - \frac{3 \epsilon}{4} 
\nonumber\\
&&-2 \epsilon L_{3M}(p,\mu)\Bigr],\\
&& L_{3M}(p,\mu) = \int_{0}^{1}dx \int_{0}^{1} dy (1-y)\times \nonumber\\
&&ln\Bigl[\frac{p^{2}}{\mu^{2}}y(1-y) - \Bigl(1 -y +\frac{y}{x(1-x)}\Bigr)\Bigr].
\end{eqnarray}
\end{subequations}
\par As pointed out in the computation of the graphs from $\Gamma^{(4)}$ concerning the finiteness of the 
integral $L_{M}(p,\mu)$ appearing there,  the integral $L_{3M}(p,\mu)$ resulting in the computation of the present vertex part is finite. The same remarks are valid for their counterparts $L_{E}(p,\mu)$ and $L_{3E}(p,\mu)$ in Euclidean space. We will discuss this topic in Appendix B. 
\par The quantity we are really interested in is the combination required in the diagrammatic expansion of 
$\Gamma^{(2)}(p)$ at two-loop order, namely
\begin{eqnarray}\label{A17}
&& -i \Bigl[\parbox{10mm}{\includegraphics[scale=0.9]{fig6.eps}} - \parbox{10mm}{\includegraphics[scale=0.9]{fig6.eps}}_{p=0}\Bigr]= \frac{(N+2)u_{0}^{2} p^{2}}{144 \epsilon}[1+\frac{\epsilon}{4} \nonumber\\
&& - 2 \epsilon L_{3M}(p,\mu)].
\end{eqnarray} 
\par The three-loop diagram of the two-point vertex function is given 
by $ -u_{0}^{3} \mu^{ 3 \epsilon} \frac{(N+2)(N+8)}{108} I_{5}$, where 
  \begin{eqnarray}\label{A18}
&I_{5}(p,\mu) = \int \frac{d^{d} q_{1} d^{d} q_{2} d^ {d} q_{3}}{(q_{1}^{2} - \mu^{2})(q_{2}^{2} - \mu^{2})(q_{3}^2 - \mu^ {2})[(q_{1} + q_{2} + p)^{2} - \mu^{2}]}\nonumber\\
& \times \; \frac{1}{[(q_{1} + q_{3} + p)^{2} - \mu^{2}]}.
\end{eqnarray}
\par This integral can be cast in the following form
\begin{subequations}\label{A19}
\begin{eqnarray}
&I_{5}(p) = \frac{2}{(3d-10)} \Bigl[\mu^{2} (C_{1M}(p,\mu) + 4 C_{2M}(p,\mu)) \nonumber\\
& \qquad - 2 D_{M}(p,\mu)\Bigr],\label{A19a}\\
&C_{1M}(p,\mu) = \int \frac{d^{d} q_{1} d^{d} q_{2} d^ {d} q_{3}}{(q_{1}^{2} - \mu^{2})^{2}(q_{2}^{2} - \mu^{2})(q_{3}^2 - \mu^ {2})[(q_{1} + q_{2} + p)^{2} - \mu^{2}]}\nonumber\\
& \times \; \frac{1}{[(q_{1} + q_{3} + p)^{2} - \mu^{2}]},\label{A19b}\\
&C_{2M}(p,\mu) = \int \frac{d^{d} q_{1} d^{d} q_{2} d^ {d} q_{3}}{(q_{1}^{2} - \mu^{2})(q_{2}^{2} - \mu^{2})^{2} (q_{3}^2 - \mu^ {2})[(q_{1} + q_{2} + p)^{2} - \mu^{2}]}\nonumber\\
& \times \; \frac{1}{[(q_{1} + q_{3} + p)^{2} - \mu^{2}]},\label{A19c}\\
&D_{M}(p,\mu) =  \int  \frac{d^{d} q_{1} d^{d} q_{2} d^{d} q_{3} p.(q_{1} + q_{2} + p)}{(q_{1}^{2} - \mu^{2}) (q_{2}^{2} - \mu^{2})(q_{3}^{2} - \mu^{2})} \nonumber\\
&\qquad \frac{1}{[(q_{1} + q_{2} + p)^{2} - \mu^{2}]^{2}[(q_{1} + q_{3} + p)^{2} - \mu^{2}]}.\label{A19d}
\end{eqnarray}
\end{subequations}
Consider  $C_{1M}$. Taking into account that this integral has $I_{2M}^{2}(q_{1}+p, \mu)$ as a subdiagram of the integration over the loop momenta $q_{1}$,  we find
\begin{subequations}
\begin{eqnarray}\label{A20}
&& C_{1M} (p,\mu) = f_{1M}(\epsilon) \int_{0}^{1}dx  \int_{0}^{1}dy \int \frac{d^{d}q_{1}}{[q_{1}^{2} - \mu^{2}]^{2}}\nonumber\\
&& \times \;\;\frac{1}{[(q_{1}+p)^{2}x(1-x) - \mu^{2}]^{\frac{\epsilon}{2}}}\nonumber\\
&& \times \;\; \frac{1}{[(q_{1}+p)^{2}y(1-y) - \mu^{2}]^{\frac{\epsilon}{2}}},\\
&& f_{1M}(\epsilon)= \frac{-1}{\epsilon^{2}}\Bigl[1 - \epsilon + \frac{1}{4} \epsilon^{2} + \frac{1}{2} \psi^{'}(1) \epsilon^{2} \Bigr].
\end{eqnarray}
\end{subequations}
Now, use the formula \cite{Yndu}
\begin{eqnarray}\label{A21}
&&\frac{1}{A^{\alpha}B^{\beta}C^{\delta}}= \frac{\Gamma(\alpha+\beta+\gamma}{\Gamma(\alpha)\Gamma(\beta)\Gamma(\gamma)} \int_{0}^{1}dz z \nonumber\\
&&\int_{0}^{1}\frac{dw u_{1}^{\alpha -1}u_{2}^{\beta -1}u_{3}^{\gamma -1}}{[u_{1}A+u_{2}B+u_{3}C]^{\alpha+\beta+\gamma}},
\end{eqnarray}
where $u_{1}=zw$, $u_{2}=z(1-w)$ and $u_{3}=1-z$.   We then find
\begin{eqnarray}\label{A22}
& C_{1M} (p,\mu) = \frac{f_{1M}(\epsilon) \Gamma(2+\epsilon)}{[\Gamma(\frac{\epsilon}{2})]^{2}} \int_{0}^{1} dx [x(1-x)]^{-\frac{\epsilon}{2}} \int_{0}^{1} dy \nonumber\\
& [y(1-y)]^{-\frac{\epsilon}{2}} \int_{0}^{1} dz  z^{1+\frac{\epsilon}{2}} (1-z)^{-1+\frac{\epsilon}{2}} 
\int_{0}^{1} dw (1-w)^{\frac{\epsilon}{2} -1} \int d^{d}q_{1} \nonumber\\
& \frac{1}{\Bigl[q_{1}^{2} + (2q_{1}.p + p^{2})(1-zw) - \mu^{2}(zw + \frac{z(1-w)}{x(1-x)} +  \frac{(1-z)}{y(1-y)})\Bigr]^{2+\epsilon}} .
\end{eqnarray}
We compute the momentum integral over $q_{1}$ at the endpoint the singularities of the parametric equation corresponding to the values $z=w=1$, which is the implementation of the $PDT$ in practice. This has the virtue of eliminating all the dependence on the external 
momenta $p$.  The parametric integrals can be easily solved independently and we find
\begin{eqnarray}\label{A23}
& C_{1M}(p,\mu)  = \frac{-i \mu^{-3 \epsilon}}{3 \epsilon^{3}} \Bigl[ 1 - \frac{\epsilon}{2} + 
\frac{3\epsilon^{2}}{4} (1 + \psi^{'}(1)) \Bigr].
\end{eqnarray}
There was a misprint in the result of reference \cite{Leite1} in the coefficient of $\psi^{'}(1)$ which is 
now corrected.
\par Consider $C_{2M}(p,\mu)$.  The integral over $q_{3}$ produces the subdiagram $I_{2M}(q_{1}+p,\mu)$.  The difference with respect with the previous integral just calculated is that one has utilize two steps: $\it{i)}$ performing the integral over $q_{2}$ using Feynman parameters again followed by $\it{ii)}$  the computation of the resulting integral using another set of Feynman parameters to integrate over $q_{1}$. These steps result in the following intermediate expression
\begin{eqnarray}
& C_{2M} (p,\mu) = f_{2M}(\epsilon)\int_{0}^{1} dx [x(1-x)]^{-\frac{\epsilon}{2}} \int_{0}^{1} dy 
y^{-\frac{\epsilon}{2}} \nonumber\\
& (1-y)^{-1 -\frac{\epsilon}{2}} \int_{0}^{1} dz  [z(1-z)]^{\frac{\epsilon}{2}}  
\int_{0}^{1} dw (1-w)^{\frac{\epsilon}{2} -1} \int d^{d}q_{1} \nonumber\\
& \frac{1}{\Bigl[q_{1}^{2} + (2q_{1}.p + p^{2})(1-zw)- \mu^{2}(zw + \frac{z(1-w)}{x(1-x)} +  \frac{(1-z)}{y(1-y)})\Bigr]^{2+\epsilon}},\label{A24}\\
& f_{2M}(\epsilon)= \frac{-1}{2\epsilon} \Bigl[1 -\epsilon + \frac{\epsilon^{2}}{2}(\frac{1}{2} + \psi^{'}(1))\Bigr].\label{A25}
\end{eqnarray}
\par Now, set $z=w=1$ inside the loop integral to get rid of the $p$ dependence of this object.  Note that although $w=1$ corresponds to the endpoint singularity of this parametric integral,  the value $z=1$ is {\it not} the endpoint singularity of the integral over $z$.  This argument can be applied to any loop order, utilizing a specific perturbative structure that excludes tadpole graphs. When focusing on a certain diagram of higher loop order after the "partial-$p$" operation, one typically ends up with different coefficients of $\mu^{2}$ for the given integral. This is shown explicitly for two and three-loop graphs. These coefficients are composed of homotopically different integrals, whose external momentum dependence is eliminated by the same choice of values in the parametric integrals, as explained above. The bottom line is that all terms proportional to the bare mass will cancel out due to the specific combination required to eliminate tadpoles.To obtain $C_{2M}$, we use the parametric dissociation transform and find
\begin{eqnarray}\label{A26}
& C_{2M}(p,\mu)  = \frac{i \mu^{-3 \epsilon}}{3 \epsilon^{3}} \Bigl[ 1 - \frac{3\epsilon}{2} + 
\frac{\epsilon^{2}}{4} (7 + 3\psi^{'}(1)) \Bigr].
\end{eqnarray}
Consider the integral $D_{M}(p,\mu)$.  Since $2D_{M}(p,\mu) = -\frac{p}{2}.\frac{\partial I_{5}}{\partial p}$, it is easy to show that
\begin{eqnarray}\label{A27}
&2D_{M}(p,\mu) = \frac{-i p^{2} \mu^{-3 \epsilon}}{6 \epsilon^{2}}\Bigl[1-\epsilon - 3\epsilon L_{3M}(p,\mu) \Bigr].
\end{eqnarray}
The combination we need in the particular perturbative expansion is given by the diagrammatic expression
\begin{eqnarray}\label{A28}
&& -i \Bigl[\parbox{10mm}{\includegraphics[scale=0.9]{fig7.eps}} - \parbox{10mm}{\includegraphics[scale=0.9]{fig7.eps}}_{p=0}\Bigr]= \frac{-(N+2)(N+8)u_{0}^{3} p^{2}}{648 \epsilon^{2}}[1+\frac{\epsilon}{2} \nonumber\\
&& - 3 \epsilon L_{3M}(p,\mu)].
\end{eqnarray} 

\section{Feynman graphs in Euclidean space}
\par  We will now highlight the differences between the previous discussion in Minkowski space and the current one to be concise.  Our starting point is the formula \cite {Amit}
\begin{eqnarray}\label{B1}
& \int \frac{d^{d} q}{[q^{2}+2kq+ \mu^{2}]^{\alpha}} = \frac{\Gamma(\frac{d}{2}) \Gamma(\alpha - \frac{d}{2})}{2 \Gamma(\alpha)} (\mu^{2} - k^{2})^{\frac{d}{2} -\alpha} .
\end{eqnarray}
\par The one-loop contribution to $\Gamma^{(4)}$ can be written as $\frac{u_{0}^{2} \mu^{2 \epsilon}(N+8)}{18} I_{2}$, which is given by the expression:  
\begin{eqnarray}\label{B2}
& \parbox{10mm}{\includegraphics[scale=0.9]{fig10.eps}}\quad = \frac{u_{0}^{2} \mu^{2 \epsilon}(N+8)}{18} \int \frac{d^{d} q}{[q^{2}+ \mu^{2}][(q+P)^{2} + \mu^{2}]}.
\end{eqnarray}
Its $\epsilon$-expansion is
\begin{eqnarray}\label{B3}
&& \parbox{10mm}{\includegraphics[scale=0.9]{fig10.eps}}= \frac{u_{0}^{2} \mu^{\epsilon}(N+8)}{18 \epsilon} \Bigl[1 - \frac{\epsilon}{2}  -\frac{\epsilon}{2} L_{E}(P,\mu)\Bigr],
\end{eqnarray}
where $(L_{E} \equiv )L_{E}(P,\mu) = \int_{0}^{1} dx ln\Bigl[\frac{P^{2}}{\mu^{2}} x(1-x) + 1 \Bigr]$ is finite.  
\par The trivial two-loop can be expressed in terms of $I_{2}$ as $ -u_{0}^{3} \mu^{3\epsilon}\frac{(N^{2}+6N+20)}{108}I_{2}^{2}$. Using the above results,  the trivial two-loop graph can be written  as
\begin{eqnarray}\label{B4}
& \parbox{10mm}{\includegraphics[scale=0.9]{fig11.eps}}\quad= - \frac{u_{0}^{3} \mu^{\epsilon}(N^{2}+6N+20)}{108\epsilon^{2}}\Bigl[1 -\epsilon -\epsilon L_{E}\Bigr].
\end{eqnarray}
\par The nontrivial diagram of the two-loop contribution for the four-point vertex part is  
$ - u_{0}^{3} \mu^{3 \epsilon} \frac{5N+22}{54} I_{4}$, which reads
\begin{eqnarray}\label{B5}
&& \parbox{10mm}{\includegraphics[scale=0.9]{fig12.eps}}\quad =- \frac{u_{0}^{3} \mu^{3 \epsilon}(5N+22)}{54}  \int \frac{d^{d} q_{1} d^{d} q_{2}}
{[q_{1}^{2} + \mu^{2}][q_{2}^{2} + \mu^{2}]}\nonumber\\
&& \times \frac{1}{[(P-q_{1})^{2} + \mu^{2}][(q_{1}- q_{2} + p_{3})^{2} + \mu^{2}]}.
\end{eqnarray} 
It can be shown that
\begin{eqnarray}\label{B6}
& \parbox{10mm}{\includegraphics[scale=0.9]{fig12.eps}} =-\frac{u_{0}^{3} \mu^{\epsilon}(5N+22)}{108 \epsilon^{2}} \Bigl[ 1 - \frac {1}{2}\epsilon - \epsilon L_{E}(P,\mu)\Bigr] .
\end{eqnarray} 
\par Consider the two-point vertex part, specifically the two- and three-loop contributions.   
The two-loop diagram is represented by the product $\frac{ u_{0}^{2} \mu^{2 \epsilon}(N+2)}{18}I_{3}$,  where $I_{3}$ is the Euclidean momentum integral, namely 
\begin{eqnarray}\label{B7}
& I_{3}= 
\int \frac{d^{d} q_{1} d^{d} q_{2}}{[q_{1}^{2} + \mu^{2}][q_{2}^{2} + \mu^{2}]} \nonumber\\
& \;\;\; \times \frac{1}{[(q_{1} + q_{2} + p)^{2}) + \mu^{2}]}.
\end{eqnarray}
After using the partial-$p$ operation as explained in the main text,  the integral can be written in the form
\begin{subequations}\label{B8}
 \begin{eqnarray}
& I_{3} = \frac{-1}{(d-3)} \Bigl[3 \mu^ {2} A_{E}(p,\mu)+ B_{E}(p,\mu) \Bigr], \label{B8a}\\
&A_{E}(p,\mu) = \int  \frac{d^{d} q_{1} d^{d} q_{2}}{[q_{1}^{2} + \mu^{2}]^{2}[q_{2}^{2} + \mu^{2}][(q_{1} + q_{2} + p)^{2} + \mu^{2}]}, \label{B8b}\\
&B_{E}(p,\mu) = \int  \frac{d^{d} q_{1} d^{d}q_{2} p. (q_{1} +q_{2} +p)}{[q_{1}^{2} + \mu^{2}]
[q_{2}^{2} + \mu^{2})[(q_{1} + q_{2} + p)^{2} + \mu^{2}]^{2}}.\label{B8c}
\end{eqnarray}
\end{subequations}
\par The integrals $A_{E}$ and $B_{E}$ within the PDT technique are analogous to their Minkowski counterparts. We simply need to write down their results:
\begin{subequations}
\begin{eqnarray}\label{B9}
&& A_{E}(p,\mu) = \frac{\mu^{-2 \epsilon}}{2 \epsilon^{2}} \Bigl[1 - \frac{\epsilon}{2} + \epsilon^{2}(\frac{\pi^{2}}{12} + \frac{1}{2})\Bigr],\label{B9a}\\
&& B_{E}(p,\mu)= \frac{\mu^{-2 \epsilon} p^{2}}{8 \epsilon}\Bigl[1 - \frac{3 \epsilon}{4} - 2 \epsilon L_{3E}(p,\mu)\Bigr]. 
\label{B9b}
\end{eqnarray}
\end{subequations}
\par The diagrammatic combination we need is given by the expression
\begin{subequations}
\begin{eqnarray}\label{B10}
&& - \Bigl[\parbox{10mm}{\includegraphics[scale=0.9]{fig6.eps}} - \parbox{10mm}{\includegraphics[scale=0.9]{fig6.eps}}_{p=0}\Bigr]= \frac{(N+2)u_{0}^{2} p^{2}}{144 \epsilon}[1+\frac{\epsilon}{4} \nonumber\\
&& - 2 \epsilon L_{3E}(p,\mu)],\label{B10a}\\
&& L_{3E}(p,\mu) = \int_{0}^{1}dx \int_{0}^{1} dy (1-y)ln\Bigl[\frac{p^{2}}{\mu^{2}}y(1-y) \nonumber\\
&& + 1 -y +\frac{y}{x(1-x)} \Bigr].\label{B10b}
\end{eqnarray}
\end{subequations}
\par Let us conclude with the three-loop diagram of $\Gamma^{(2)}$.   This diagram corresponds to $ -u_{0}^{3} \mu^{ 3 \epsilon} \frac{(N+2)(N+8)}{108} I_{5}$, where 
  \begin{eqnarray}\label{B11}
&I_{5}(p,\mu) = \int \frac{d^{d} q_{1} d^{d} q_{2} d^ {d} q_{3}}{(q_{1}^{2} + \mu^{2})(q_{2}^{2} + \mu^{2})(q_{3}^2 + \mu^ {2})[(q_{1} + q_{2} + p)^{2} + \mu^{2}]}\nonumber\\
& \times \; \frac{1}{[(q_{1} + q_{3} + p)^{2} + \mu^{2}]}.
\end{eqnarray}
\par As we discussed earlier, if we apply partial-$p$ to this integral, we can simplify it.  Then, we find
\begin{subequations}\label{B12}
\begin{eqnarray}
&I_{5}(p) = \frac{-2}{(3d-10)} \Bigl[\mu^{2} (C_{1E}(p,\mu) + 4 C_{2E}(p,\mu)) \nonumber\\
& \qquad + 2 D_{E}(p,\mu)\Bigr],\label{B12a}\\
&C_{1E}(p,\mu) = \int \frac{d^{d} q_{1} d^{d} q_{2} d^ {d} q_{3}}{(q_{1}^{2} + \mu^{2})^{2}(q_{2}^{2} + \mu^{2})(q_{3}^{2} + \mu^ {2})}\nonumber\\
& \times \; \frac{1}{[(q_{1} + q_{2} + p)^{2} + \mu^{2}][(q_{1} + q_{3} + p)^{2} + \mu^{2}]},\label{B12b}\\
&C_{2E}(p,\mu) = \int \frac{d^{d} q_{1} d^{d} q_{2} d^ {d} q_{3}}{(q_{1}^{2} + \mu^{2})(q_{2}^{2} + \mu^{2})^{2} (q_{3}^{2} + \mu^ {2})}\nonumber\\
& \times \; \frac{1}{[(q_{1} + q_{2} + p)^{2} + \mu^{2}][(q_{1} + q_{3} + p)^{2} + \mu^{2}]},\label{B12c}\\
&D_{E}(p,\mu) =  \int  \frac{d^{d} q_{1} d^{d} q_{2} d^{d} q_{3} p.(q_{1} + q_{2} + p)}{(q_{1}^{2} + \mu^{2}) (q_{2}^{2} + \mu^{2})(q_{3}^{2} + \mu^{2})} \nonumber\\
&\qquad \frac{1}{[(q_{1} + q_{2} + p)^{2} + \mu^{2}]^{2}[(q_{1} + q_{3} + p)^{2} + \mu^{2}]}.\label{B12d}
\end{eqnarray}
\end{subequations}
\par The objects $C_{1E}$ and $C_{2E}$ are computed using the parametric dissociation transform, which renders them independent of external momentum. Refer to the main text for the explicit intermediate steps. Collecting the results, we obtain:
\begin{subequations}
 \begin{eqnarray}\label{B13}
& C_{1E}(p,\mu)  = \frac{ \mu^{-3 \epsilon}}{3 \epsilon^{3}} \Bigl[ 1 - \frac{\epsilon}{2} + 
\frac{3\epsilon^{2}}{4} (1 + \psi^{'}(1)) \Bigr],\label{B13a}\\
& C_{2E}(p,\mu)  = \frac{- \mu^{-3 \epsilon}}{3 \epsilon^{3}} \Bigl[ 1 - \frac{3 \epsilon}{2} + 
\frac{7 \epsilon^{2}}{4} + \frac{3 \epsilon^{2}}{4}\nonumber\\
& \times \psi^{'}(1)) \Bigr],\label{B13b}\\
&2D_{E}(p,\mu) = \frac{ p^{2} \mu^{-3 \epsilon}}{6 \epsilon^{2}}\Bigl[1-\epsilon - 3\epsilon L_{3E}(p,\mu) \Bigr].\label{B13c}
\end{eqnarray}
\end{subequations}
First ,note that a misprint appeared in the last term of $C_{1}(p,\mu)(=C_{1E}(p,\mu))$ in ref. \cite{Leite1}, which is now corrected.  It led to an error in the expression $C_{1}(p,\mu)+ 4C_{2}(p,\mu)(=C_{1E}(p,\mu)+4C_{2E}(p,\mu))$. The correct expression is $C_{1}(p,\mu)+ 4C_{2}(p,\mu)(=C_{1E}(p,\mu)+4C_{2E}(p,\mu))= \frac{- \mu^{-3 \epsilon}}{\epsilon^{3}} \Bigl[ 1 - \frac{11 \epsilon}{6} + 
\frac{25 \epsilon^{2}}{12} + \frac{3 \epsilon^{2}}{4}\psi^{'}(1)) \Bigr]$.
\par What is required in the perturbative expansion is actually the graph subtracted from its value at zero external momentum. The result is
\begin{eqnarray}\label{B14}
&& - \Bigl[\parbox{10mm}{\includegraphics[scale=0.9]{fig7.eps}} - \parbox{10mm}{\includegraphics[scale=0.9]{fig7.eps}}_{p=0}\Bigr] = u_{0}^{3}  \frac{p^{2}(N+2)(N+8)}{648 \epsilon^{2}}\Bigl[1\nonumber\\
&&+\frac{\epsilon}{2} - 3\epsilon L_{3E}(p,\mu) \Bigr].
\end{eqnarray}
\par In the vertex part $\Gamma^{(2,1)}$, with both metric signatures, the diagrams are proportional to the appropriate integrals $I_{2}$ and $I_{4}$ of the vertex part $\Gamma^{(4)}$. The exact expressions corresponding to these integrals can be found in the main text.
\par When we compare the solution to the diagrams in Minkowski with their Euclidean counterparts, we notice that the results in both cases are quite similar, except for a global factor in the diagrams. Furthermore, the final expressions for the graphs with Euclidean signature involve the substitution of $L_{E}(P,\mu)$ and $L_{3E}(P,\mu)$ with $L_{M}(P,\mu)$ and $L_{3M}(P,\mu)$, respectively, with Minkowski signature and the same coefficients. This simplifies the comparison of renormalization by minimal subtraction with both signatures.

\end{document}